\documentstyle[psfig]{mn} 
 
\newif\ifAMStwofonts 
 
\ifoldfss 
  \ifCUPmtlplainloaded \else 
    \NewTextAlphabet{textbfit} {cmbxti10} {} 
    \NewTextAlphabet{textbfss} {cmssbx10} {} 
    \NewMathAlphabet{mathbfit} {cmbxti10} {} 
    \NewMathAlphabet{mathbfss} {cmssbx10} {} 
  \fi 
  \ifAMStwofonts 
    \ifCUPmtlplainloaded \else 
      \NewSymbolFont{upmath} {eurm10} 
      \NewSymbolFont{AMSa} {msam10} 
      \NewMathSymbol{\upi}     {0}{upmath}{19} 
      \NewMathSymbol{\umu}     {0}{upmath}{16} 
      \NewMathSymbol{\upartial}{0}{upmath}{40} 
      \NewMathSymbol{\leqslant}{3}{AMSa}{36} 
      \NewMathSymbol{\geqslant}{3}{AMSa}{3E}

    \fi 
  \fi 
\fi 
 
\ifnfssone 
  \newmathalphabet{\mathit} 
  \addtoversion{normal}{\mathit}{cmr}{m}{it} 
  \addtoversion{bold}{\mathit}{cmr}{bx}{it} 
  \def\textbfit{\protect\txtbfit}
  \def\textbfss{\protect\txtbfss}
 
\long\def\txtbfit#1{{\fontfamily{cmr}\fontseries{bx}\fontshape{it
}%
    \selectfont #1}}
 
\long\def\txtbfss#1{{\fontfamily{cmss}\fontseries{bx}\fontshape{n
}%
    \selectfont #1}}
  \newmathalphabet{\mathbfit} 
\textbfit{..}
  \addtoversion{normal}{\mathbfit}{cmr}{bx}{it}
  \addtoversion{bold}{\mathbfit}{cmr}{bx}{it}
  \newmathalphabet{\mathbfss} 
\textbfss{..}
  \addtoversion{normal}{\mathbfss}{cmss}{bx}{n}
  \addtoversion{bold}{\mathbfss}{cmss}{bx}{n}
  \ifAMStwofonts
    \ifCUPmtlplainloaded \else
italic and
if your
      \UseAMStwoboldmath
      \makeatletter
      \new@mathgroup\upmath@group
      \define@mathgroup\mv@normal\upmath@group{eur}{m}{n}
      \define@mathgroup\mv@bold\upmath@group{eur}{b}{n}
      \edef\UPM{\hexnumber\upmath@group}
      \new@mathgroup\amsa@group
      \define@mathgroup\mv@normal\amsa@group{msa}{m}{n}
      \define@mathgroup\mv@bold\amsa@group{msa}{m}{n}
      \edef\AMSa{\hexnumber\amsa@group}
      \makeatother
      \mathchardef\upi="0\UPM19
      \mathchardef\umu="0\UPM16
      \mathchardef\upartial="0\UPM40
      \mathchardef\leqslant="3\AMSa36
      \mathchardef\geqslant="3\AMSa3E
    \fi
  \fi
\fi 

\ifnfsstwo
  \def\textbfit{\protect\txtbfit}
  \def\textbfss{\protect\txtbfss}
 
\long\def\txtbfit#1{{\fontfamily{cmr}\fontseries{bx}\fontshape{it
}%
    \selectfont #1}}
 
\long\def\txtbfss#1{{\fontfamily{cmss}\fontseries{bx}\fontshape{n
}%
    \selectfont #1}}
  \DeclareMathAlphabet{\mathbfit}{OT1}{cmr}{bx}{it}
  \SetMathAlphabet\mathbfit{bold}{OT1}{cmr}{bx}{it}
  \DeclareMathAlphabet{\mathbfss}{OT1}{cmss}{bx}{n}
  \SetMathAlphabet\mathbfss{bold}{OT1}{cmss}{bx}{n}
  \ifAMStwofonts
    \ifCUPmtlplainloaded \else
      \DeclareSymbolFont{UPM}{U}{eur}{m}{n}
      \SetSymbolFont{UPM}{bold}{U}{eur}{b}{n}
      \DeclareSymbolFont{AMSa}{U}{msa}{m}{n}
      \DeclareMathSymbol{\upi}{0}{UPM}{"19}
      \DeclareMathSymbol{\umu}{0}{UPM}{"16}
      \DeclareMathSymbol{\upartial}{0}{UPM}{"40}
      \DeclareMathSymbol{\leqslant}{3}{AMSa}{"36}
      \DeclareMathSymbol{\geqslant}{3}{AMSa}{"3E}
    \fi
  \fi
\fi 

\ifCUPmtlplainloaded \else
  \ifAMStwofonts \else 
    \def\upi{\pi}
    \def\umu{\mu}
    \def\upartial{\partial}
  \fi
\fi

\title[NAT formation]{Cluster-subcluster mergers and the formation of narrow-angle tailed radio sources}
\author[M. Bliton et al.]
        {M.~Bliton,$^{1,2}$\thanks{mbliton@hades.physics.missouri.edu} E.~Rizza,$^{1,2}$ J.O.~Burns,$^{1,2,3}$ F.N.~Owen,$^4$
        and M.J.~Ledlow$^{1,5}$\\
	$~1$ Department of Astronomy, New Mexico State University, Las Cruces, NM USA 88003\\
        $~2$ Department of Physics \& Astronomy, University of Misouri, Columbia, MO USA 65211\\
	$~3$ Office of Research, University of Missouri, Columbia, MO USA 65211\\
        $~4$ National Radio Astronomy Observatory \thanks{The National
        Radio Astronomy Observatory is a 
        facility of the U.S. National Science Foundation operated under
        cooperative agreement by Associated Universities, Inc.}, P.O. Box O,
        Socorro, NM USA 87801\\
	$~5$ Department of Physics \& Astronomy, University of New
Mexico, Albuquerque, NM USA 87131\\}

\date{Accepted 1998 May 18.
      Received 1998 April 15;
      in original form 1997 April 7}

\pagerange{\pageref{firstpage}--\pageref{lastpage}}
\pubyear{1998}

\begin{document}

\label{firstpage}

\maketitle

\begin{abstract}
We have examined the {\it ROSAT} PSPC X-ray properties of a sample of
15 Abell clusters containing 23 narrow-angle tailed (NAT) radio
galaxies.  We find that clusters with NATs show a significantly higher
level of substructure than a similar sample of radio-quiet clusters,
indicating that NAT radio sources are preferentially located in
dynamically complex systems.  Also, the velocity distribution of the
NAT galaxies is similar to that of other cluster members; these
velocities are inadequate for producing the ram pressure necessary to
bend the radio jets.  We therefore propose a new model for NAT
formation, in which NATs are associated with dynamically complex
clusters undergoing merger events.  The {\bf U}-shaped NAT morphology
is produced in part by the merger-induced bulk motion of the ICM
bending the jets.
\end{abstract}


\section{Introduction}
The interaction between extragalactic radio sources and the rich
cluster environment in which they are often found is not well
understood.  Radio galaxies in the same or similar types of clusters
are often found to have widely varying radio morphologies.  One
possible explanation for the variety of radio structures and
characteristics is the ``weather'' within the intracluster medium
(ICM). As the radio jets propagate through the hot gas, variations in
the ICM density and velocity may account for the complex radio
structures that are observed (Burns 1996).  The interactions between
radio sources and the ICM are discussed in the Burns et al. (1994a)
study of radio galaxies in Abell clusters using {\it Einstein} IPC
data.  They found that 75 per cent of all radio galaxies $>$2 arcmin
in diameter have X-ray substructure within 5 arcmin of the host galaxy
position, presumably due to overdense regions in the ICM.
Additionally, radio galaxies tend to be more concentrated toward the
centres of rich clusters, where the ICM is the densest, in comparison
with the optical distribution of cluster galaxies (Ledlow \& Owen
1995).

Two specific examples of the interaction between radio sources and the
ICM are wide-angle tailed (WAT) and steep spectrum radio sources.
WATs are radio sources located predominantly at the centre of rich
clusters in D or cD galaxies, which appear to be bent by $\rm \approx
1000 \ km \ s^{-1}$ bulk motions of the ICM resulting from
cluster-subcluster mergers (G\'omez et al. 1997; Roettiger, Burns \&
Loken 1996).  Steep spectrum sources are amorphous objects found in
the centre of cooling flow clusters (Burns 1990). The cooling flow
disrupts the jets (Loken et al. 1993; Burns et al. 1997), and the
thermal pressure reduces the adiabatic expansion of the radio plasma.
The steep spectrum is thereby produced from loss of the plasma's high
energy electrons through inverse Compton and synchrotron processes
(Slee \& Reynolds 1984).

Perhaps the most striking example of the interaction between the ICM
and radio sources are the head-tail, or narrow-angle tailed (NAT)
radio sources (O'Dea \& Owen 1985).  These sources have radio jets
that are bent at extreme angles--up to $\rm 90^{\circ}$ from their
original orientation.  The standard explanation for this radio
morphology is that the jets are curved by ram pressure from the high
velocity host galaxy moving through the dense ICM.  This jet bending
can place important constraints on both the jet dynamics and the ICM
(Burns \& Owen 1980).  Jones \& Owen (1979, JO) and Begelman, Rees \&
Blandford (1979, BRB) used detailed maps of the prototypical NAT
galaxy NGC 1265 to test models for ram pressure-induced jet bending.
Numerical simulations tested the validity of these models (Balsara \&
Norman 1992), showing that both the JO and BRB models produce the
observed NAT morphology and properties (Balsara, priv. comm.).  An
intriguing characteristic of narrow-angle tailed radio sources is that
they have also been observed in poor clusters of galaxies (Doe et
al. 1995).  Such environments often have central ICM densities an
order of magnitude lower than in rich clusters.  Lower ICM densities
require either lower radio pressures or substantially higher galaxy
velocities in order to produce the ram pressure capable of bending the
jets (Venkatesan et al. 1994).

A number of studies examined the relationship between NAT radio
sources and the cluster environments in which they reside.  Feretti,
Perola \& Fanti (1992) determined that NAT radio galaxies in a sample
of Abell clusters appeared to be underpressured with respect to the
surrounding ICM.  More recently, Edge \& R\"ottgering (1995, ER)
examined the coincidence of excess X-ray emission and tailed radio
sources in clusters.  This research was motivated in part by the Burns
et al. (1994a) result described above.  Using higher resolution {\it
ROSAT} PSPC images, ER detected compact, apparently unresolved, X-ray
emission associated with NATs and concluded that this emission was
caused by an AGN.  However, they did not examine the relationship
between the NATs and the larger-scale ICM.  Specifically, does the
local ICM play a role in shaping NAT galaxies, similar to the current
model for WATs?

Two recent events have made this an excellent time to examine this
hypothesis.  First, all {\it ROSAT} PSPC data are now available in the
public archives, and second, the completion of the 20 cm VLA Abell
cluster survey (see Owen \& Ledlow 1997; Ledlow \& Owen 1996, and
references therein) has produced a large sample of NAT galaxies in
rich clusters.  Together, these databases allow us to study the
interaction of NAT galaxies with the rich Abell cluster environment
for a representative sample of clusters.

The paper is organized as follows: In \S 2 we describe the samples of
NAT and radio-quiet clusters used in this paper. In \S 3 we discuss
the data analysis, including the X-ray substructure, velocity, and
alignment tests.  \S 4 contains our discussion of the results in the
context of a new model for the formation of NAT radio sources.  In \S
5 we list our conclusions.  Throughout this paper, we assume
$H_{\circ} \rm = 75 \ km \ s^{-1} \ Mpc^{-1}$ and $q_{\circ} = 0.5$.

\section{The samples}
Two samples were utilized in this analysis: a radio-loud sample of
clusters containing NATs, and a radio-quiet sample of clusters
containing no detected radio sources.  To define our first sample, we
began with the complete, 20 cm survey of radio galaxies in Abell
clusters (Ledlow \& Owen 1995).  This sample consists of all radio
sources in Abell clusters of $\ z < 0.09$ with 1.4 GHz flux density
($S_{\rm 1.4 GHz}$) $>$ 10 mJy and within 0.3 Abell cluster radii
$(A_{\rm c})$ or $\approx$ 660 kpc.  Using the high resolution radio
maps from Owen \& Ledlow (1997), we examined the morphology of the
radio galaxies.  All radio sources showing head-tail or tadpole
morphology, or resolved jets curved backward into a {\bf U}-shape were
classified as NATs.  Questionable identifications were removed from
the sample yielding a total of 54 NATs in Abell clusters.  Since the
goal of this study was to examine the interaction between narrow-angle
tailed radio galaxies and the ICM, we cross referenced our NAT sample
with the {\it ROSAT} PSPC archive.  We retained only clusters with
X-ray images that have signal-to-noise (S/N) $\rm >$ 20 within a 0.5
Mpc radius aperture (see \S 3.1 for further details on S/N), and were
within the inner support ring of the PSPC.  This final cut reduced the
sample to 23 NAT galaxies in 15 host Abell clusters.  Although the
sample is no longer complete because of the requirement for pointed
PSPC observations, we feel the resultant sample is representative of
NAT galaxies in rich clusters.  The final sample is summarized in
Table 1 and X-ray/radio overlays are shown in Fig. 1.

The radio-quiet sample was also constructed from the PSPC archives.
This sample includes all clusters from the VLA Abell cluster survey
having no known radio sources with $S_{\rm 1.4 GHz} > 10 \ \rm mJy$,
and with redshift less than 0.09.  We then chose clusters from the
PSPC archives using the same criteria as the NAT clusters.  We also
added one cluster (A2244) at $z$=0.0968 to increase the sample size to
13 radio-quiet clusters, which are summarized in Table 2.  We chose a
radio-quiet control sample (as opposed to a sample having any radio
sources that were not NATs) for several reasons.  Firstly,
morphological classification of radio sources can be difficult and
highly subjective, and although this must be done for the NAT sample,
we wanted to lessen this effect as much as possible.  Secondly, for
the X-ray analysis, there was no longer a need to remove the
contaminating X-rays from the AGN at the centres of the radio sources
for our control sample.  Thirdly, small, tailed radio sources may be
misclassified as compact sources due to resolution limits.  Lastly,
the NAT morphology is only seen in low power radio sources, where the
radio jets are less ``stiff'' and therefore more susceptible to ram
pressure induced jet bending.  However, higher power radio sources
with stiffer jets may not display the morphological effects of ram
pressure, even if it is present.  In Table 3 we show a comparison of
the cluster properties for the two samples.  A few illustrative
examples of these radio-quiet clusters are shown in Fig. 2.

\begin{table*}
 \centering
 \begin{minipage}{140mm}
  \caption{NAT cluster properties}
  \begin{tabular}{@{}cccccccc@{}}
   Abell \# & IAU Name & Cluster $z$\footnote{Struble \& Rood (1991)} &
ROR \# & Exposure (s) & Rood--Sastry & Richness class\footnote{Abell, Corwin \&
Olowin (1989)}(\#)\footnote{Number of cluster members (Abell, Corwin \& Olowin 1989)} & $\sigma_{\rm v}$$^a$ $\rm (km \ s^{-1}$)\\[10pt] 
85   & 0039-095A & 0.0556 & 800250 & 10240 & cD & 1(59) & 749 \\
     & 0039-097  &        &        &       &    &       & \\
119  & 0053-015  & 0.0444 & 800251 & 14700 & C  & 1(69) & 778 \\
     & 0053-016  &        &        &       &    &       & \\
194  & 0123-016A & 0.0180 & 800316 & 24500 & L  & 0(37) & 440 \\
496  & 0431-134  & 0.0327 & 800024 & 8600  & cD & 1(50) & 741 \\
514  & 0445-205  & 0.0734 & 800278 & 17500 & F  & 1(78) & \\
     & 0446-205  &        &        &       &    &       & \\
754  & 0906-095  & 0.0534 & 800232 & 6200  & cD & 2(92) & 1048 \\
1314 & 1131+493  & 0.0338 & 800392 & 2800  & C  & 0(44) & 664 \\
     & 1132+492  &        &        &       &    &       & \\
1367 & 1142+198  & 0.0214 & 800153 & 18100 & F  & 2(117)& 822 \\
     & 1141+202B &        &        &       &    &       & \\
1656 & 1256+281  & 0.0231 & 800005 & 20400 & B  & 2(106)& 880 \\
1775 & 1339+266B & 0.0724 & 701068 & 12900 & B  & 2(92) & 1594 \\
1795 & 1346+268B & 0.0622 & 800055 & 25000 & cD & 2(115)& 896 \\
2142 & 1556+274  & 0.0896 & 800233 & 4800  & B  & 2(89) & 1241 \\
2255 & 1712+640  & 0.0808 & 800512 & 14600 & C  & 2(102)& 1221 \\
     & 1712+641  &        &        &       &    &       & \\
2256 & 1705+786  & 0.0581 & 100110 & 16600 & B  & 2(88) & 1270 \\
     & 1706+786  &        &        &       &    &       & \\
     & 1706+787  &        &        &       &    &       & \\
2589 & 2321+164  & 0.0416 & 800526 & 7100  & cD & 0(40) & 500 \\
\end{tabular}
\end{minipage}
\end{table*}

\begin{table*}
 \centering
 \begin{minipage}{140mm}
  \caption{Radio-quiet sample properties}
  \begin{tabular}{@{}ccccccc@{}}
   Abell \# & Cluster $z$\footnote{Struble \& Rood (1991)} & ROR \# &
Exposure (s) & Rood--Sastry & Richness class\footnote{Abell, Corwin \&
Olowin (1989)}(\#)\footnote{Number of cluster members (Abell, Corwin
\& Olowin 1989)} & $\sigma_{\rm v}$$^a$ $\rm (km s^{-1}$) \\[10pt] 
 539 & 0.0291 & 800255 & 9600  &  F & 1(50) & 701 \\
 548 & 0.0414 & 800041 & 3800  &  F & 1(79) & 872 \\
 644 & 0.0704 & 800379 & 11200 & cD & 0(42) & \\
 744 & 0.0729 & 700385 & 6100  &  B & 0(42) & 812 \\
1139 & 0.0383 & 701198 & 4320  &  I & 0(39) & \\
1377 & 0.0514 & 800106 & 4900  &  B & 1(59) & 488 \\
1651 & 0.0845 & 800353 & 10600 & cD & 1(70) & 965 \\
1691 & 0.0722 & 800295 & 10600 & cD & 1(64) & \\
1750 & 0.0855 & 800553 & 23400 & F  & 0(40) & 778 \\
1837 & 0.0376 & 800243 & 15700 & cD & 1(50) & 218\footnote{Zabludoff
et al. 1993} \\
2107 & 0.0421 & 800509 & 8300  & cD & 1(51) & 536 \\
2244 & 0.0968 & 800265 & 2965  & cD & 2(89) & 1240 \\
2670 & 0.0761 & 800420 & 54700 & cD & 3(142) & 881 \\
\end{tabular}
\end{minipage}
\end{table*}

\begin{table*}
 \centering
 \begin{minipage}{140mm}
  \caption{Comparison of samples}
  \begin{tabular}{@{}ccccccc@{}}
\multicolumn{1}{c}{} &
\multicolumn{3}{c}{Mean} &
\multicolumn{3}{c}{Variance$^a$} \\
Cluster property & NAT & RQ & Sig. & NAT & RQ & Sig. \\
 & & & per cent & & & per cent \\[10pt]
$z$ & 0.051 & 0.062 & 22 & 0.00055 & 0.00053 & 98 \\
Rood--Sastry type & 2.3 & 2.6 & 71 & 2.0 & 4.1 & 19 \\
Richness \# & 79 & 64 & 18 & 740 & 881 & 75 \\
X-ray luminosity ($\times10^{44}$) & 1.5 & 1.0 & 44 & 4.9 & 1.3 & 3 \\
\end{tabular}
\medskip

$^a$The variances of redshift and
richness seem extremely high due to the effects of having flat
distributions (i.e. large $\sigma$), and that the means are
significantly different from unity.  This causes the variances ($\sigma^2$) to appear very large or very small.
\end{minipage}
\end{table*}

\begin{figure*}
\centerline{\hbox{
\psfig{figure=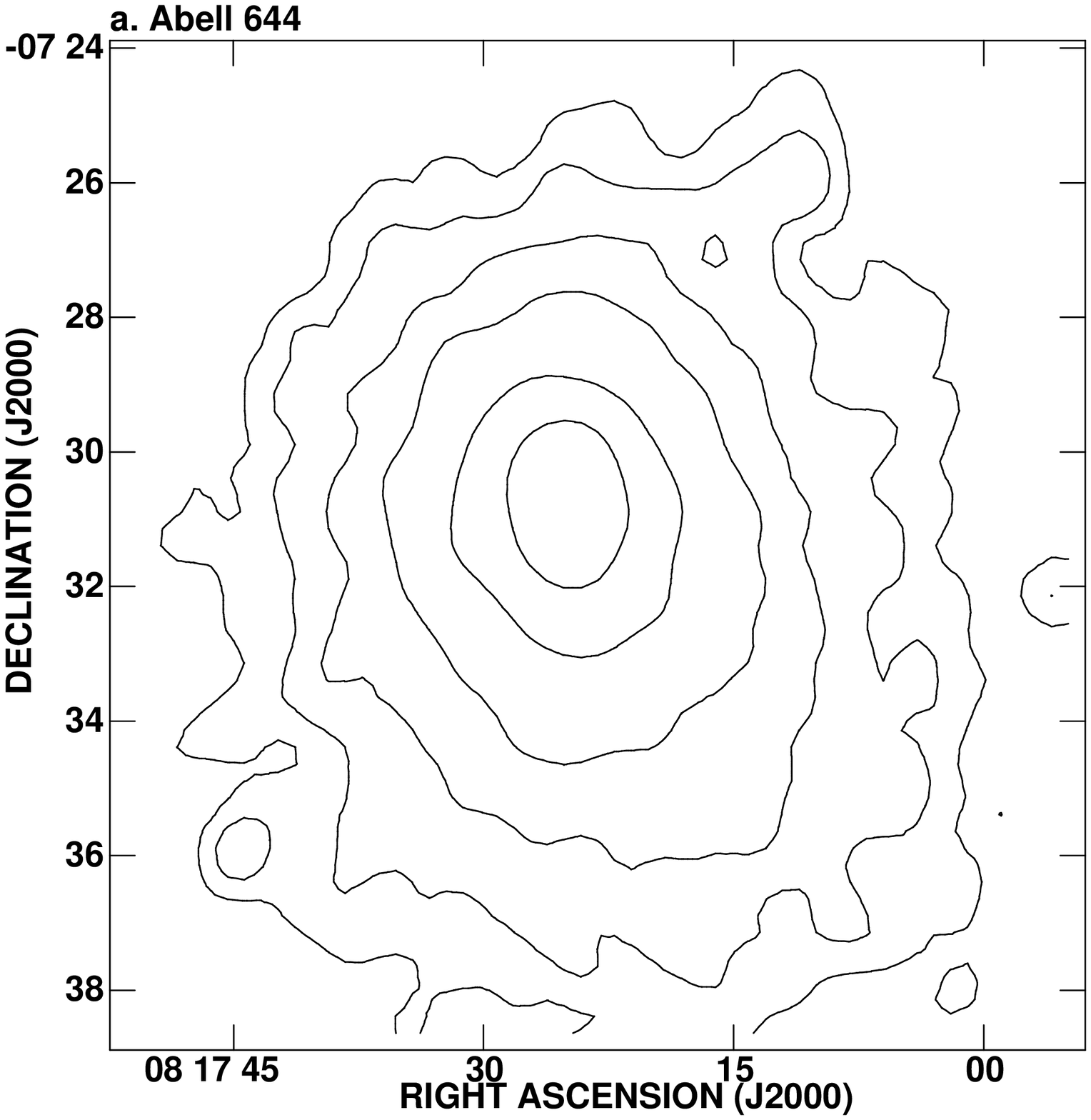,height=3in}
\psfig{figure=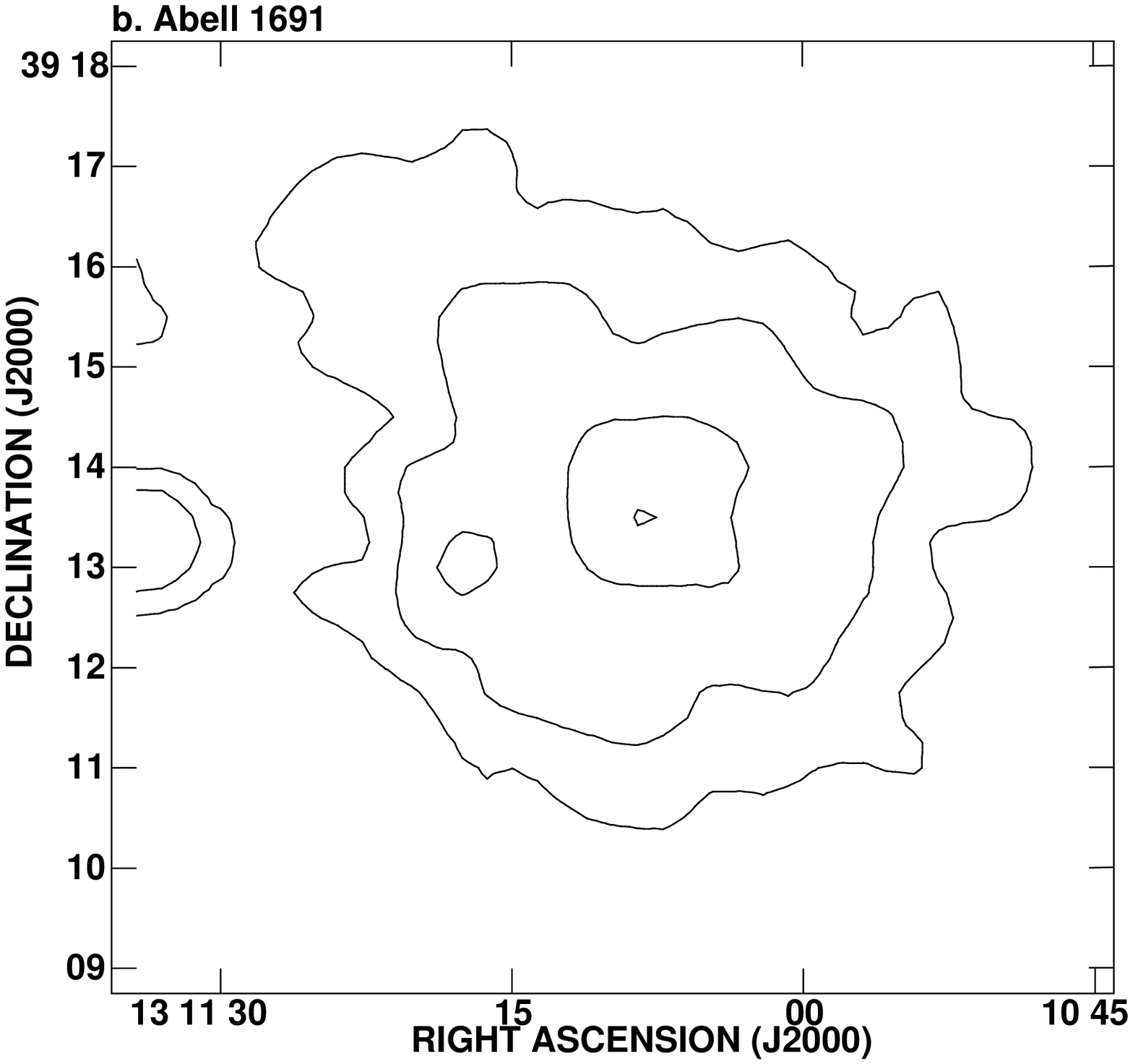,height=3in}}}
\centerline{\psfig{figure=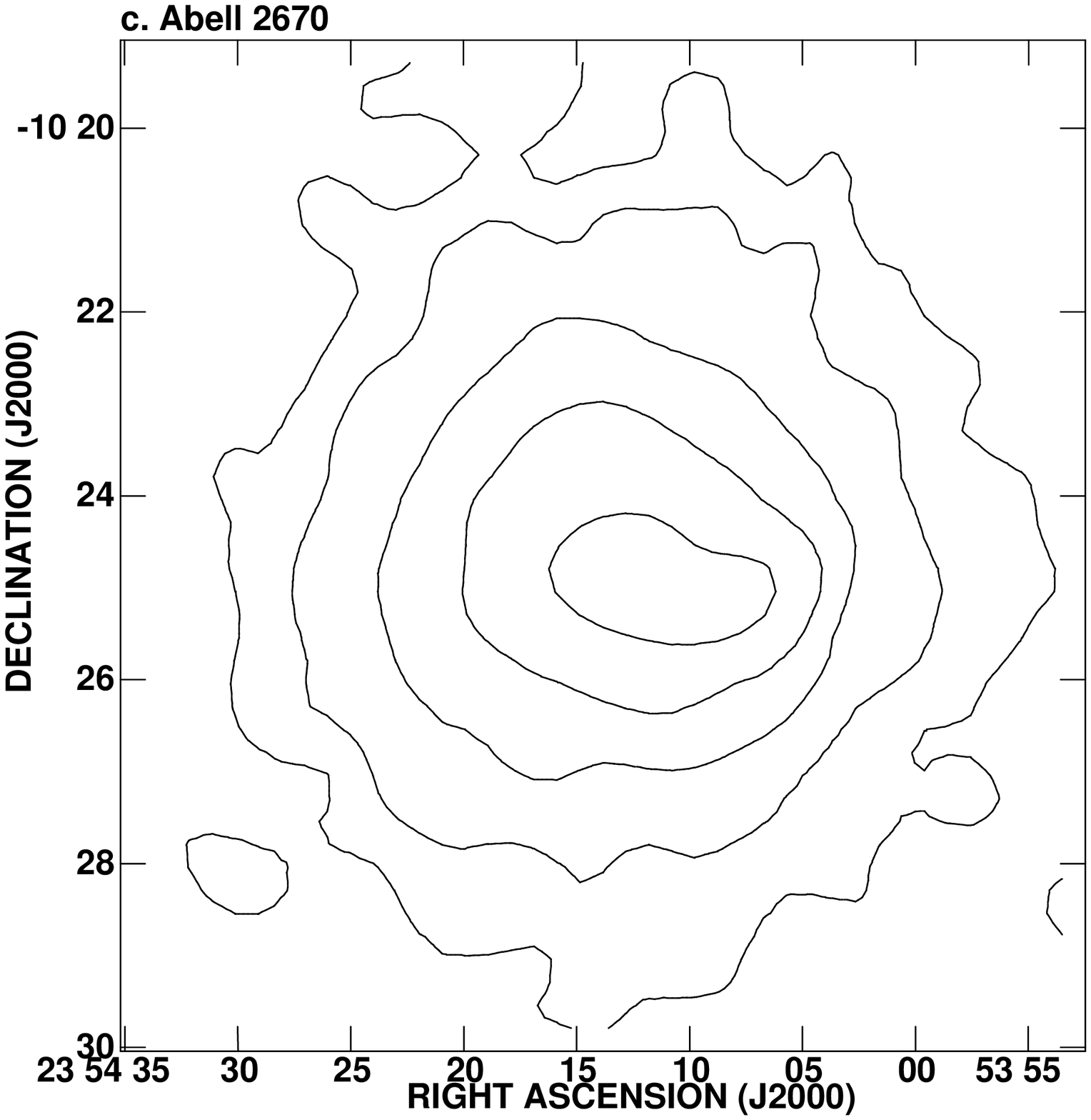,height=3in}}
\label{fig2}
\caption{{\it ROSAT} PSPC X-ray surface brightness contours of three
radio quiet clusters, (a) Abell 644, (b) Abell 1691, and (c) Abell
2670.  For all maps, X-ray surface brightness contours are at
$3\sigma, \ 5\sigma, \ 10\sigma, \ 20\sigma, \ 50\sigma, \ \& \
100\sigma.$}
\end{figure*}

In order to quantify the similarity between the two samples, we
performed statistical tests on the mean and variance of the
distributions of redshift, Rood--Sastry type (Rood \& Sastry 1971),
richness, and X-ray luminosity.  First, we used the Student's
T-statistic (Havlicek \& Crain 1988) to compare the mean of each
distribution.  This statistic tests the validity of the null
hypothesis that the two distributions have the same mean.  Values of
significance $<$5 per cent require one to reject the null hypothesis.
The significance levels for this test are shown in Table 3; note all
significance levels are above 5 per cent.  We used the F-test
(Havlicek \& Crain 1988) to determine if the distributions have
significantly different variances in the cluster parameters.  Similar
to the Student's T-statistic, the F-test examines the null hypothesis
that the two samples have the same variance.  The significance values
for this test (Table 3) again show that for most cluster properties,
the null hypothesis cannot be rejected. The one exception is in X-ray
luminosity where the variances of the two samples are similar at only
the 3 per cent level.  However, since the mean luminosities are not
significantly different and all other cluster properties are similar,
we feel that these samples are adequate for comparison.

\section{Data analysis}
The X-ray analysis used the PSPC hard-band (0.4--2.4 keV) count maps
and exposure maps from the {\it ROSAT} archive at the Goddard Space
Flight Center.  Average background values for the sample were
determined within the inner support ring of the PSPC and subtracted
from the count maps. The only exceptions were A754, A1367, and A1656,
where the cluster emission itself completely filled the inner ring
region.  The count maps were divided by the exposure maps to produce
surface brightness images in units of $\rm counts \ s^{-1} \
pixel^{-1}$, where each pixel is $\rm 15 \times 15 \ arcsec^2$.  All
X-ray data were smoothed with a 45 arcsec FWHM (32.6 kpc for $z$=0.04)
Gaussian filter.  The {\sevensize AIPS} software package was used for
aligning and resizing the images as well as producing overlays of the
radio/X-ray images.  In Fig. 1, we show radio greyscales overlaid on
to X-ray surface brightness contours for the NAT sample.  Comments on
individual clusters and NATs are given in the Appendix (\S 6).

The luminosity for each cluster was determined by extracting a
circular region of diameter 0.5 Mpc, centred on the peak of the X-ray
emission.  The extracted region was modeled with the best-fitting
Raymond \& Smith thermal plasma model (Raymond \& Smith 1977), with 30
per cent solar metal abundance, using the {\sevensize XSPEC} X-ray
spectral fitting software.  This model was then used to compute the
0.4--2.4 keV X-ray luminosity for each cluster.

\subsection{Substructure}
For a fully virialized cluster gravitational potential, one would
expect to observe a smoothly varying X-ray surface brightness profile.
However, this is not what we see in many of the NAT clusters (Fig. 1).
For example, Abell 85 (Fig. 1a) possesses a clear elongation to the
south with a clump of extended emission in the vicinity of a NAT.
Abell 514 (Fig. 1e) and Abell 1314 (Fig. 1h) exhibit clumpy,
elongated, and highly non-circular X-ray structures.  The X-ray peak
of Abell 754 (Fig. 1g) is markedly offset from the centre of the X-ray
emission.  These clear examples of cluster substructure manifest
themselves in significant radial variations of the centroid ($
R_{\circ}$) of the X-ray image.  Mohr, Fabricant \& Geller (1993) have
argued that a centroid shift in the X-ray surface brightness profile
of a cluster is a robust measure of substructure.  Since the X-ray
surface brightness distribution of a cluster is a direct measure of
the ICM density distribution, a centroid shift in X-ray emission
indicates a non-smooth density distribution, which in turn indicates
an dynamically complex cluster.  This direct link between centroid
shift and a cluster's dynamical state makes this test a strong measure
of cluster substructure.

\begin{figure}
\psfig{figure=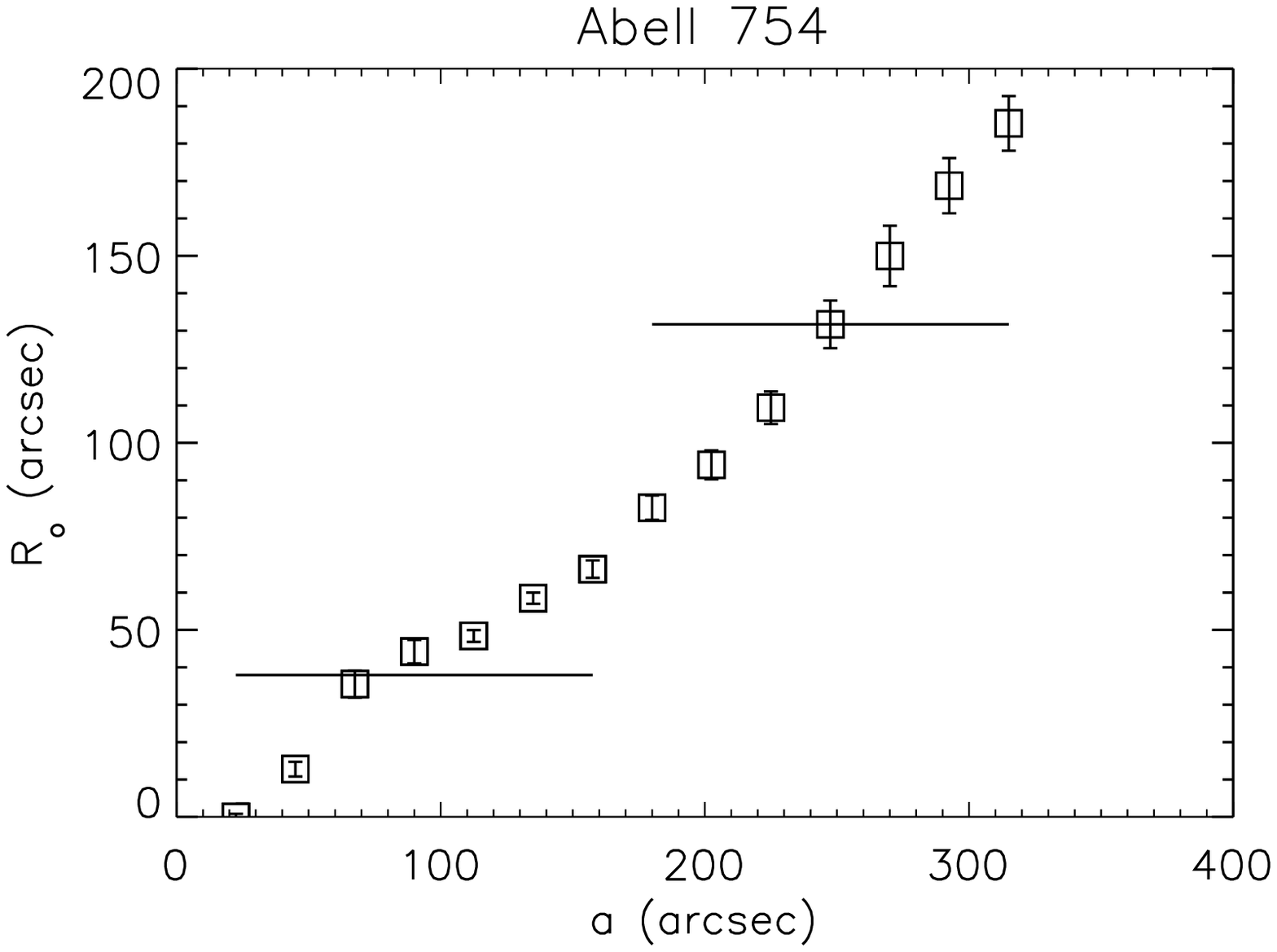,width=3.3in}
\psfig{figure=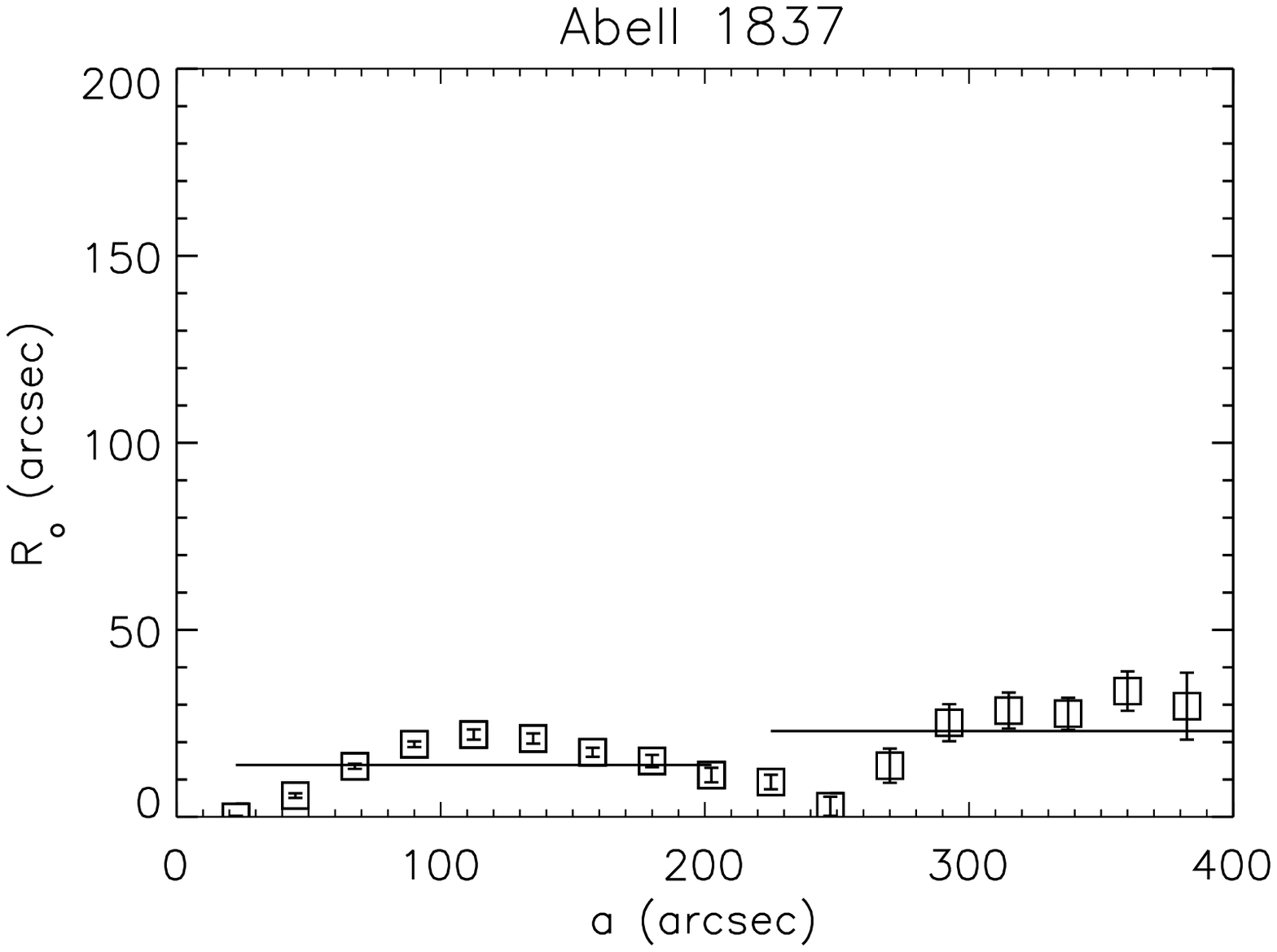,width=3.3in}
\label{fig3}
\caption{Radial variation of $R_{\circ}$ for
(a) Abell 754 and (b) Abell 1837.  The horizontal lines mark the regions
 0--150 kpc, and 150--300 kpc.}
\end{figure}

In an effort to quantify cluster substructure, elliptical isophotes
were fit to the X-ray maps using the {\sevensize IRAF STSDAS} task
ELLIPSE, based on the algorithm by Jedrzejewski (1987).  The initial
centroid was set to the X-ray peak of the cluster emission, and all
contaminating point sources were masked out.  The ELLIPSE task
performs an iterative fit to the emission weighted centroid of the
X-ray distribution as a function of semimajor axis, $a$.  To increase
the signal-to-noise, we used smoothed images for the isophotal fits.
We therefore kept the step in semimajor axis, $a$, to be at least 1.5
pixels (2$a$ = 45 arcsec) in order to avoid correlated errors in
adjacent radial bins.

The ELLIPSE task produces a radial profile of $x$ and $y$ (RA \&
Dec.) centroid coordinates.  These two components were converted into
a radial vector, $ R_{\circ}$, at each bin.  For each cluster, the
radial profiles were further divided into two portions--one region
from the cluster centre to a semimajor axis of 150 kpc, and a second
region from 150 to 300 kpc.  The values of $ R_{\circ}$ were averaged
for the two regions, and these two average values were differenced to
obtain $ \Delta R_{\circ}$.  To illustrate this process we show the
radial variation of $ R_{\circ}$ for clusters with large (A754) and
small (A1837) centroid shifts in Fig. 3.

Histograms showing the distributions of $ \Delta R_{\circ}$ for both
the NAT and radio-quiet cluster samples are shown in Fig. 4.
Examination of the figures show that NAT clusters exhibit
preferentially larger values of $ \Delta R_{\circ}$ than the
radio-quiet clusters.  To determine the statistical significance of
this difference, the student's T-test was performed.  The means of the
$\Delta R_{\circ}$ distributions revealed the samples to be similar at
a 0.3 per cent level of significance, indicating that the two
distributions have significantly different means.  A similar result is
obtained when using the Kolmogrov--Smirnov (KS) test (1.7 per cent
significance level), which is sensitive to the shapes of the
distributions.

While performing this analysis, we were concerned about the use of the
ELLIPSE routine in {\sevensize IRAF} to determine the centroid shift.
Are the solutions generated by ELLIPSE stable, and how large are the
errors associated with this task?  Do the errors depend on the
signal-to-noise of the data?  We have attempted to answer these
questions by generating errors for the centroid shifts using a
bootstrap resampling algorithm.  First, we generated 1000 realizations
of every cluster count map by reassigning values to each pixel based
on the Poisson probability distribution, with the original pixel value
as the mean of that distribution.  Therefore, each resampled pixel
could have a large range of possible values, but was most likely to
have the value of the original data.  We then ran the identical
ELLIPSE task on all 1000 realizations, giving us a range of centroid
shifts.  From this distribution of $\Delta R_{\circ}$, we used the
best-fitting Gaussian function to determine the mean and standard
deviation of the centroid shift for each cluster, which are shown in
Table 4.  As expected, the largest standard deviations occur in
clusters with the lowest S/N.  The smallest generally occur in
clusters possessing strong cooling flows, where the S/N is high, and
the core substructure low.

To quantify whether the two distributions were significantly different
given the bootstrap errors, we calculated the means of the $\Delta
R_{\circ}$ values for each sample, weighting the means by their
standard deviations.  We also computed the errors on the means from
the individual errors on $ \Delta R_{\circ}$.  The means of the two
distributions, shown at the bottom of Table 4, differ by 3.6$\sigma$,
yielding a probability of $>$99.9 per cent that the two distributions
have different means.  Therefore, we again conclude that clusters
containing NATs show a higher degree of substructure than similar
radio-quiet clusters.

\begin{figure}
\psfig{figure=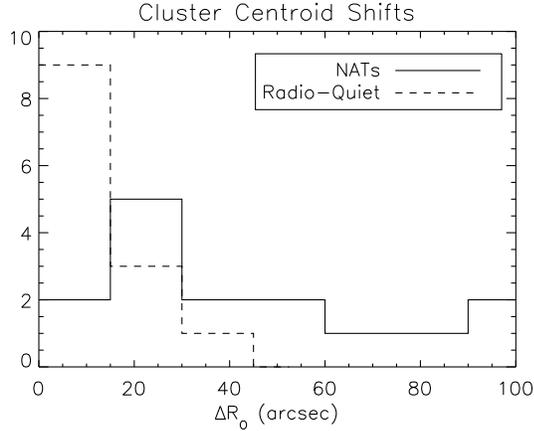,width=3.3in}
\label{fig4}
\caption{ Histograms of $\Delta R_{\circ}$ 
values for the NAT and radio-quiet samples.  The NAT sample shows a
significantly broader distribution of $\Delta R_{\circ}$ values than the
radio-quiet clusters.}
\end{figure}

\begin{table*}
 \centering
 \begin{minipage}{140mm}
  \caption{Results of bootstrap analysis}
  \begin{tabular}{@{}rrrrrrrr@{}}
\multicolumn{4}{c}{NAT Clusters} &
\multicolumn{4}{c}{Radio-quiet Clusters}\\
Abell \# & $\Delta R_{\rm actual}$ & $\Delta R_{\rm centre}$ & $
\sigma_{\rm \Delta R}$ & Abell \# & $\Delta R_{\rm actual}$ & $\Delta
R_{\rm centre}$ & $\sigma_{\rm \Delta R}$ \\
 & (arcsec) & (arcsec) & & & (arcsec) & (arcsec) & \\[10pt]
  85 & 24 & 24 &  1.8 &  539 & 11 & 12 &  6.8 \\
 119 & 15 & 11 & 15.6 &  548 &  2 & 27 & 35.5 \\
 194 & 19 & 13 & 43.7 &  644 & 24 & 24 &  2.2 \\
 496 & 28 & 28 &  3.2 &  744 &  8 & 11 &  7.7 \\
 514 & 52 & 54 & 11.5 & 1139 & 36 & 40 & 22.2 \\
 754 & 94 & 93 & 11.4 & 1377 & 20 & 27 & 15.0 \\
1314 & 42 & 49 & 28.3 & 1651 &  6 &  7 &  1.8 \\
1367 & 61 & 55 & 25.6 & 1691 &  5 &  3 &  5.4 \\
1656 & 93 & 93 &  6.2 & 1750 &  7 &  8 &  3.8 \\
1775 & 32 & 32 &  4.1 & 1837 &  9 & 10 &  7.2 \\
1795 & 16 & 16 &  1.0 & 2107 &  6 &  6 &  4.1 \\
2142 &  3 &  3 &  1.1 & 2244 &  4 &  4 &  2.4 \\
2255 & 57 & 59 &  6.1 & 2670 & 18 & 17 &  3.7 \\
2256 & 90 & 89 &  4.8 &      &    &    &      \\
2589 &  5 &  7 &  3.3 &      &    &    &      \\
\multicolumn{4}{c}{Mean: 16} &
\multicolumn{4}{c}{Mean: 11}\\
\multicolumn{4}{c}{$\sigma_{\rm mean}$: 0.6} &
\multicolumn{4}{c}{$\sigma_{\rm mean}$: 1.0}\\
\end{tabular}
\end{minipage}
\end{table*}

\subsection{Velocities}
The X-ray substructure result above implies that these clusters are in
an interesting, non-relaxed state.  To pursue this further, we
examined the dynamics of the NATs with respect to the rest of the
cluster galaxies.  Radio sources in general tend to have low peculiar
velocities compared to other cluster members (Owen, Ledlow \& Keel
1995), yet in the standard model for narrow-angle tailed radio source
formation, NATs should be associated with high velocity galaxies.
Large peculiar motions of NAT galaxies are required to produce the ram
pressure necessary to bend the radio tails into the observed {\bf U}
shapes.

For this velocity analysis, we returned to the complete sample of 54
NATs.  In order to compare galaxies in different clusters, all galaxy
velocities were normalized to their cluster's velocity dispersion, by
using the parameter $U = (v_{\circ} - v_{\rm c})/\sigma$, where $
v_{\circ}$ is the galaxy velocity, $ v_{\rm c}$ is the mean cluster
velocity, and $ \sigma$ is the cluster velocity dispersion (Owen et
al. 1995).  The NATs in the sample were compared to radio-quiet
cluster galaxies as well as other radio sources.  The general radio
source sample consisted of all radio galaxies, excluding NATs, in the
Ledlow \& Owen (1995) sample for which there were velocity data
available (Owen et al. 1995).  Since the NATs were removed from this
sample, most of these radio sources were morphologically linear, and
not sharply bent.  These sources are expected to have velocities
significantly lower than NAT galaxies, if the standard model for NAT
formation is correct. Since $\sigma$ must be used in this analysis,
only NATs in clusters with published velocity dispersions were
included, reducing the sample size to 31.  To simulate typical cluster
galaxies, we generated a Gaussian distribution of velocities using
Monte Carlo sampling.  The simulated distribution was compared with
the NAT sample, as well as the other radio sources, to check the
similarity of the distributions.  A histogram comparing the three
distributions is shown in Fig. 5.

Using the student's T-test, we find the mean of the NAT galaxy
distribution is statistically similar to both the simulated cluster
galaxies and the other radio sources.  This result is contrary to the
standard model of NAT formation, which requires the host galaxy to
have a large peculiar motion ($\sim$ 1000 km $\rm s^{-1}$) in order
to produce the ram pressure necessary to bend the radio jets.  One
would therefore expect a much broader distribution in $U$ (typical $U
\approx$ 1.0) for NAT galaxies compared to other radio sources.  The
observed distribution of NAT velocities is clearly incapable of
producing the needed ram pressure for jet bending.

There is a potential projection effect which could skew the above
analysis.  In order to classify radio galaxies as NATs, the tails must
be projected away from the observer's line of sight.  Any NAT with its
radio tail parallel to the line-of-sight will not be classified as a
NAT. If the direction of NAT tails is due primarily to the motion of
the host galaxy, then the NATs with the highest line-of-sight
velocities will not be included in this analysis.  Therefore, the
observed NAT distribution may be peaked toward low velocities by
excluding objects with high line-of-sight velocities.  We attempted to
determine the magnitude of this effect by calculating the change in an
expected distribution of NAT velocities once galaxies moving parallel
to the line-of-sight are removed.  For our idealized distribution of
NATs, we produced a Gaussian 3D velocity distribution centred at 1000
km $\rm s^{-1}$ larger than a cluster's central velocity.  This is
chosen since velocities on the order of 1000 km $\rm s^{-1}$ are
considered necessary for jet bending (O'Dea 1985; Eilek et al. 1984).
The Gaussian distribution had a $\sigma$ = 400 km $\rm s^{-1}$, in
order to leave only 10 per cent of the total number of NATs with
velocities less than 500 km $\rm s^{-1}$.  We then gave the NATs
random projection angles distributed uniformly about the sky, and
calculated the NATs' line-of-sight velocity distribution from their 3D
velocities and their projection angles.

To simulate the effects of not detecting NATs with their radio tails
pointing toward the observer, we removed all galaxies with 1-D
velocities $\rm < 45^{\circ}$ from the line-of-sight.  This
distribution was converted to $U$ as described above, and compared to
the observed NAT velocity distribution (Fig. 6).  Using the student's
T-test, we find that the observed NAT distribution and the simulated
NAT distribution have significantly different means.  Clearly,
projection effects cannot account for the observed NAT distribution.
Therefore, our conclusion stands that NATs have velocities incapable
of bending their radio jets via ram pressure.

\begin{figure}
\psfig{figure=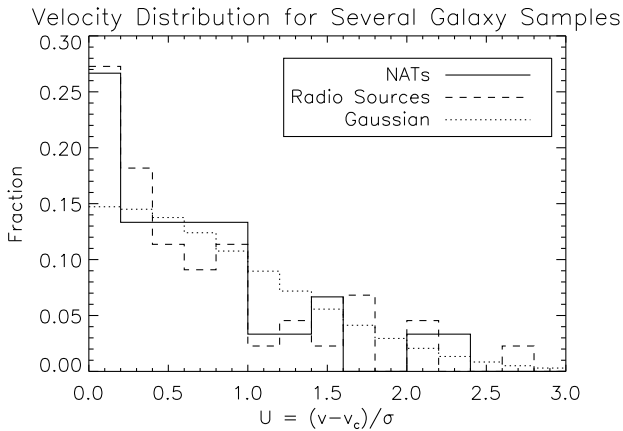,width=3.3in}
\label{fig5}
\caption{The velocity distributions for the sample of
NATs, all other radio sources, and a simulated Gaussian galaxy
cluster velocity distribution.}
\end{figure}

\begin{figure}
\psfig{figure=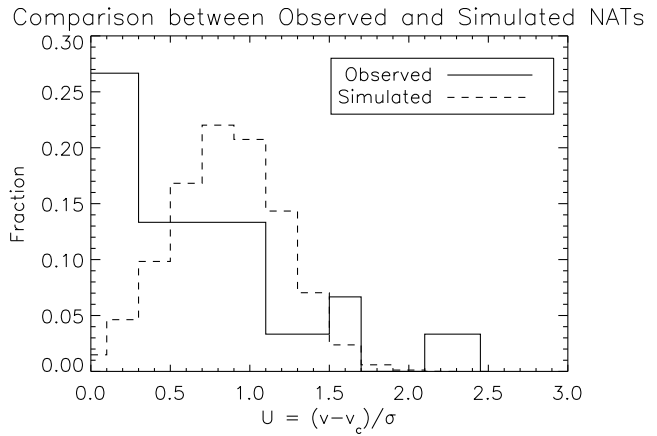,width=3.3in}
\label{fig6}
\caption{The velocity distributions for the observed NAT sample and
the simulated NAT sample.  The simulated sample does not include NATs with
velocity vectors $\rm < 45^{\circ}$ from the line-of-sight.}
\end{figure}

\subsection{Radio tail orientation}
In an effort to understand the orbits of NAT galaxies, O'Dea, Sarazin,
\& Owen (1987) examined the directions of NAT tails
with respect to the Abell cluster centres.  Assuming the NAT tails are
indicators of the direction of motion of the host galaxies, the
resulting random orientation led them to suggest that the overall
distribution of NATs is consistent with isotropic galaxy orbits.
However, only considering NATs located within 0.5 Mpc, the galaxies
exhibited a trend towards radial orbits.

We performed the same analysis as O'Dea et al. (1987) for our NAT
sample with a slight modification.  This modification was motivated by
the fact that a number of the NAT sources possess bent or curved
tails, leading to ambiguity in determining a single position angle
(e.g. NATs in Abell 514 and Abell 2255).  We therefore defined two
radio position angles: $\theta_{\rm I}$--the angle formed by a line
from the head of the radio emission to the point at which the tail
bends, and $ \theta_{\rm F}$--the angle formed by a line from the
point at which the tail bends to the 3$\sigma$ end of the radio tail.
For the NAT sources which did not show significant bending, the
overall tail direction was used for both $\theta_{\rm I}$ and
$\theta_{\rm F}$.

$\theta_{\rm I}$ and $\theta_{\rm F}$ were subtracted from the
position angle of the radial vector drawn from the NAT to the Abell
cluster centre, yielding $\Delta\theta_{\rm I}$ and $\Delta\theta_{\rm
F}$.  The relatively flat distribution in $\Delta\theta$ appears
consistent with random tail orientations (Fig. 7).  To test the
statistical significance of this hypothesis, we compared the two
samples with a theoretical, random distribution using the KS test.
For $\Delta \theta_{\rm I}$, the datasets are drawn from the same
distribution at the 91.4 per cent confidence level.  For $\Delta
\theta_{\rm F}$, the datasets are drawn from the same distribution at
the 59.9 per cent confidence level.  We therefore see no indication
that this NAT sample deviates from a random orientation relative to
the optical Abell cluster centre.

Using the same initial and final position angles as above, we compared
the tail orientations to a line between the NAT and the peak of the
cluster X-ray emission. This was done under the assumption that the
X-ray peak may give a better indication of the centre of the cluster
potential well.  The $\Delta \theta_{\rm I}$ and $\Delta \theta_{\rm
F}$ distributions for this analysis are shown in Fig. 8.  The KS test
was applied yielding significance levels of 74 per cent for
$\Delta\theta_{\rm I}$ and 84 per cent for $\Delta\theta_{\rm F}$ with
respect to a random distribution.  Again, the directions of the NAT
tails are consistent with random orientations in clusters.

\begin{figure}
\psfig{figure=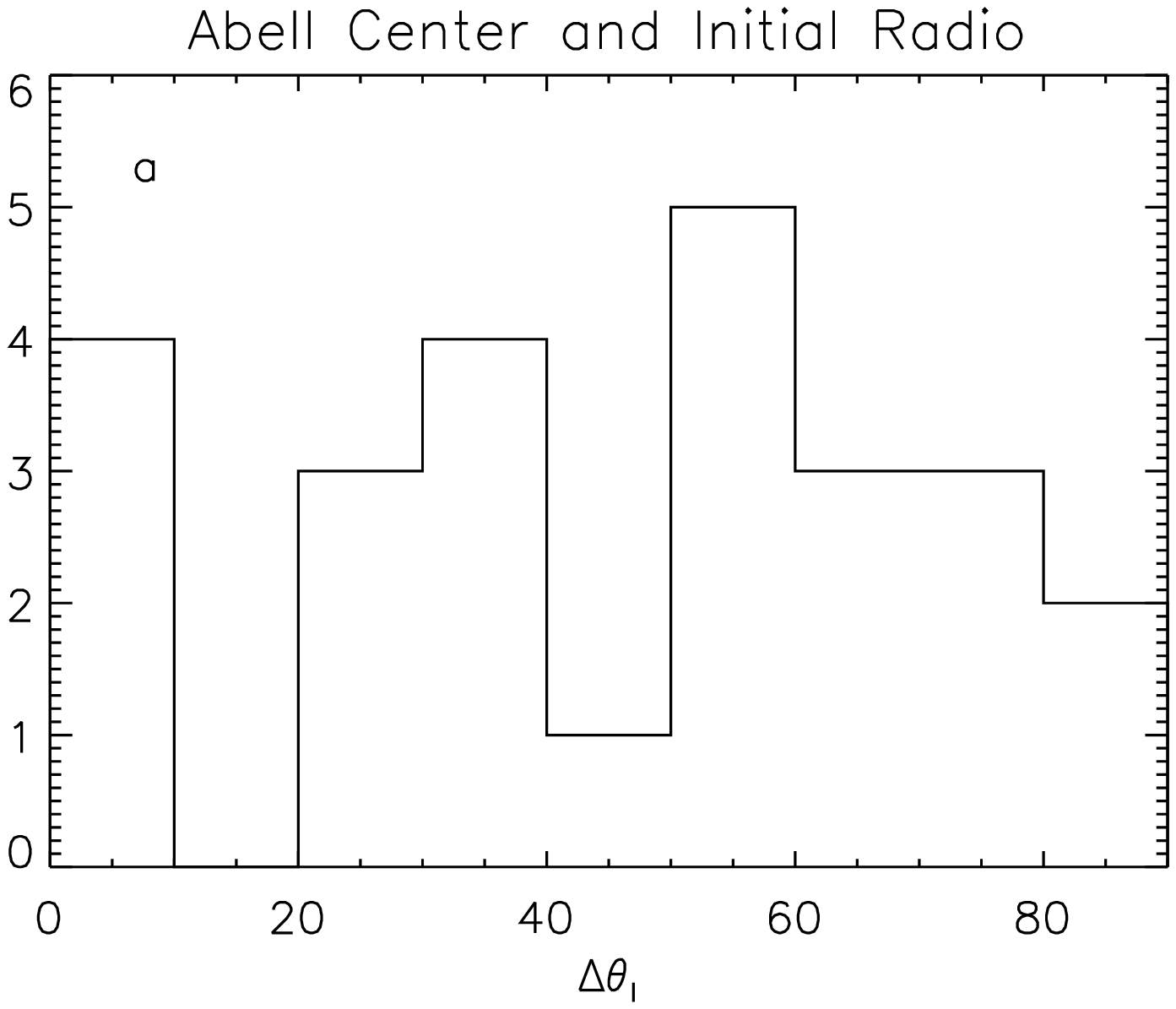,width=3.3in}
\psfig{figure=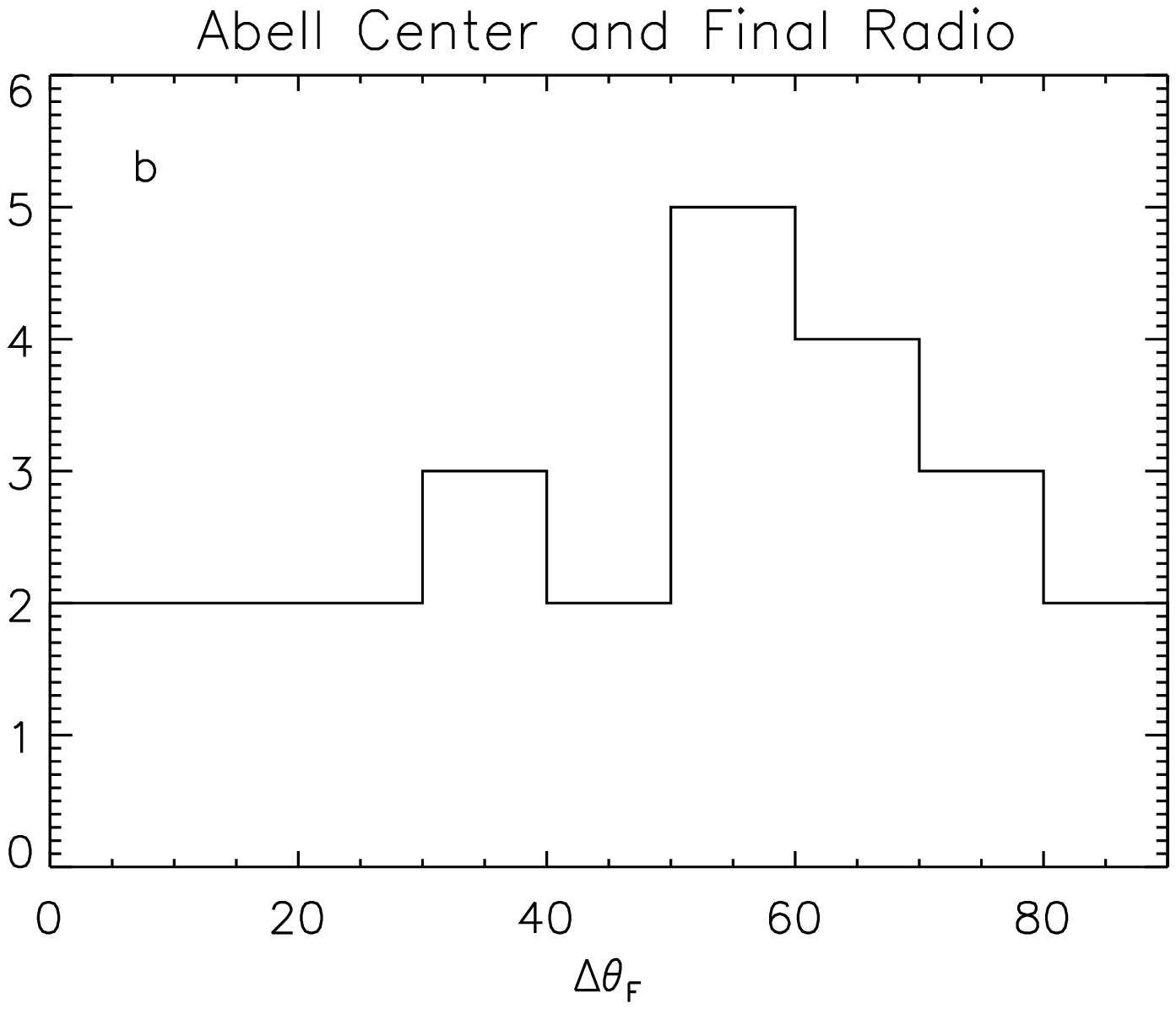,width=3.3in}
\label{fig7}
\caption{The angle between the NAT tails and the line connecting 
the NAT to the Abell cluster centre for both the (a) initial tail
direction and the (b) final tail direction.  Both distributions are
consistent with an isotropic distribution of NAT tails.}
\end{figure}

\subsection{Radio tails and X-ray substructure}
Although we found no correlation between the radio tails and the
X-ray/optical cluster centres, there is a potential alignment between
the radio tails and X-ray cluster elongations (see e.g. A119, A1656,
and A2142 in Fig. 1).  To quantify this potential result, we examined
the orientations of the radio tails with respect to the regions of
X-ray excess or elongation. We defined two X-ray position angles:
$\theta_{\rm local}$--the position angle of an X-ray enhancement
located within 1 arcmin of a NAT radio source, and $\theta_{\rm
global}$--the position angle of the entire X-ray emitting cluster.

To measure these position angles we needed to determine regions of
significant excess throughout the entire cluster.  Due to the rapid
fall-off of surface brightness with distance from the cluster centre,
the isophotal variation analysis described in \S 3.1 is only sensitive
in the inner $\rm \sim$8 arcmin (450 kpc for z=0.05) of the clusters.
Therefore, in order to probe excess emission in the outer regions of
the cluster, as well as global cluster elongations, we used an
alternative probe for substructure detailed in G\'omez et al. (1997).
We fit circular models to the cluster emission and subtracted them
from the parent map.  We specifically chose circular models to
increase our sensitivity to elongation in the overall cluster
emission.  The residual maps were examined for regions of excess
within 1 arcmin of a NAT radio position, and these excess regions were
tested for significance using Poisson statistics.  For each region of
interest, the number of counts in the cluster image was compared to a
similar region in the model image.  The summed probability of
obtaining the number of observed counts above the model value was
calculated, and any region of 99 per cent significance or higher was
considered to be substructure.  The regions of excess were used to
calculate the position angle of the local substructure, $\theta_{\rm
local}$.  The global position angle was determined at the radius where
the X-ray emission fell to approximately 5$\sigma$ above the
background.

We examined the correlation between each of the four possible
radio-X-ray position angles by measuring the difference between the
NAT and cluster orientations.  These distributions (shown in Fig. 9)
were compared with a theoretical, random distribution similar to the
analysis described in \S 3.4.  None of the distributions deviated
significantly from a random distribution.

\begin{figure}
\psfig{figure=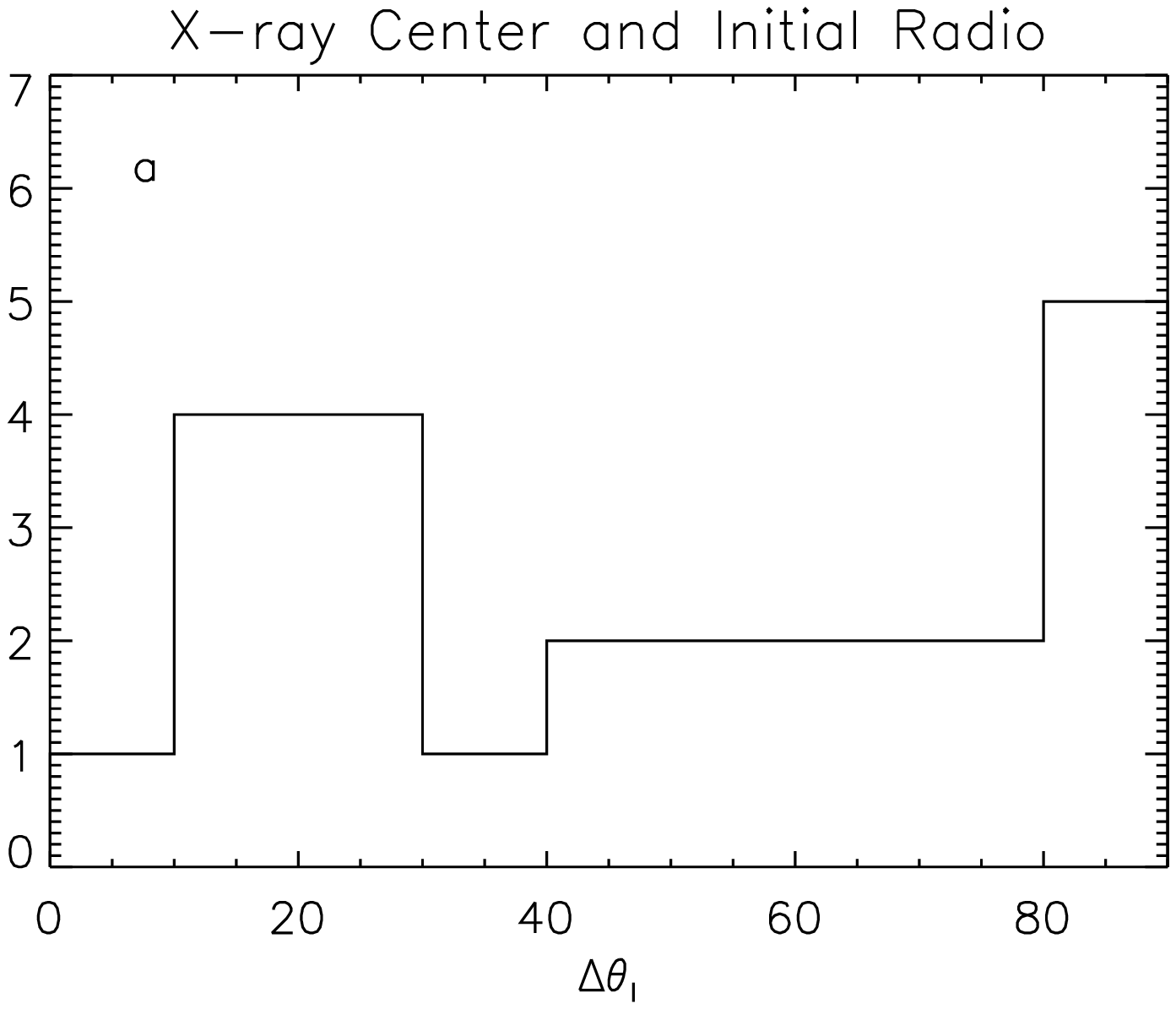,width=3.3in}
\psfig{figure=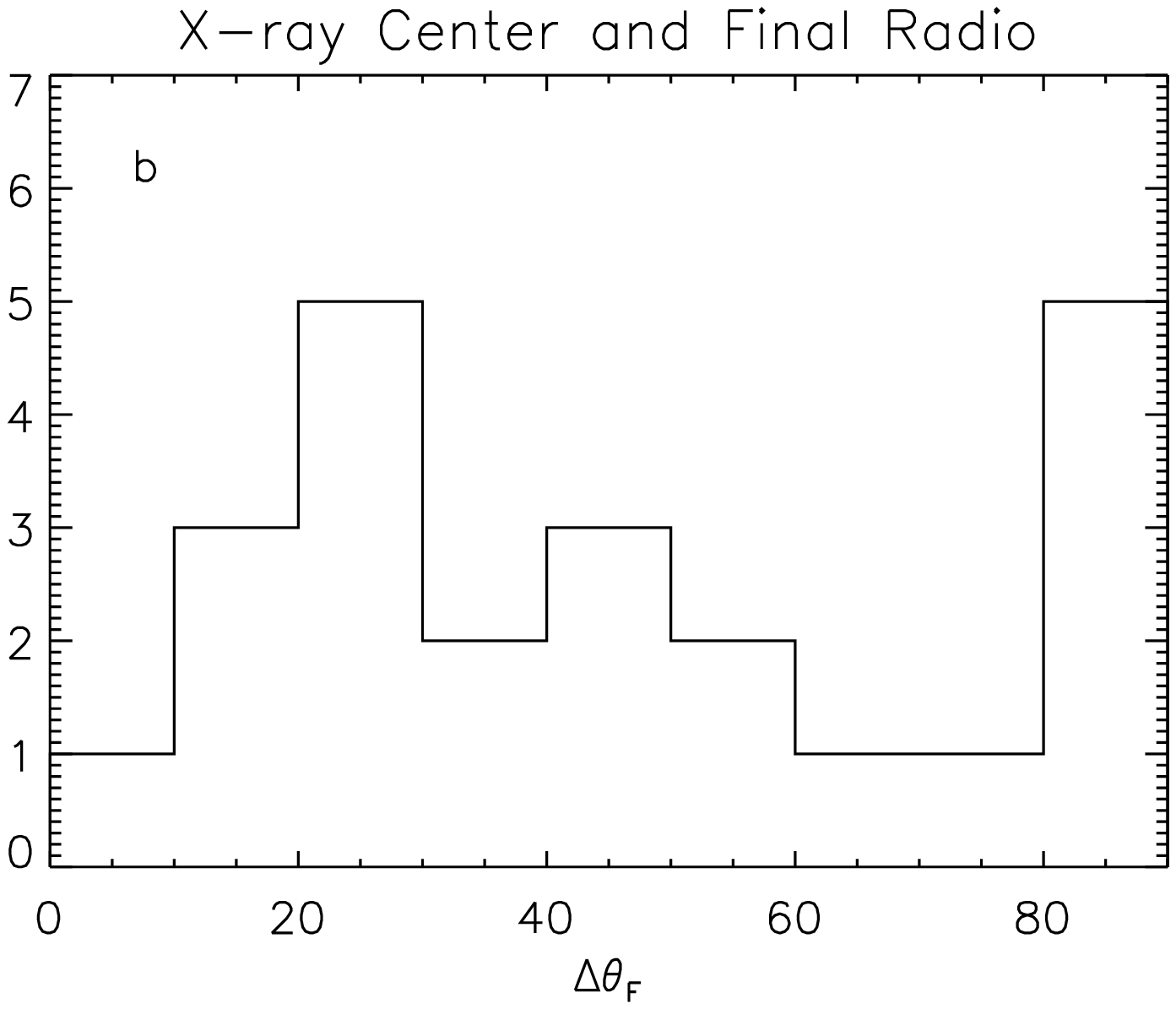,width=3.3in}
\label{fig8}
\caption{The angle between the NAT tails and the line connecting 
the NAT to the cluster X-ray peak for both the (a) initial tail
direction and the (b) final tail direction.  Both distributions are
consistent with an isotropic distribution of NAT tails.}
\end{figure}

\begin{figure*}
\centerline{\hbox{
\psfig{figure=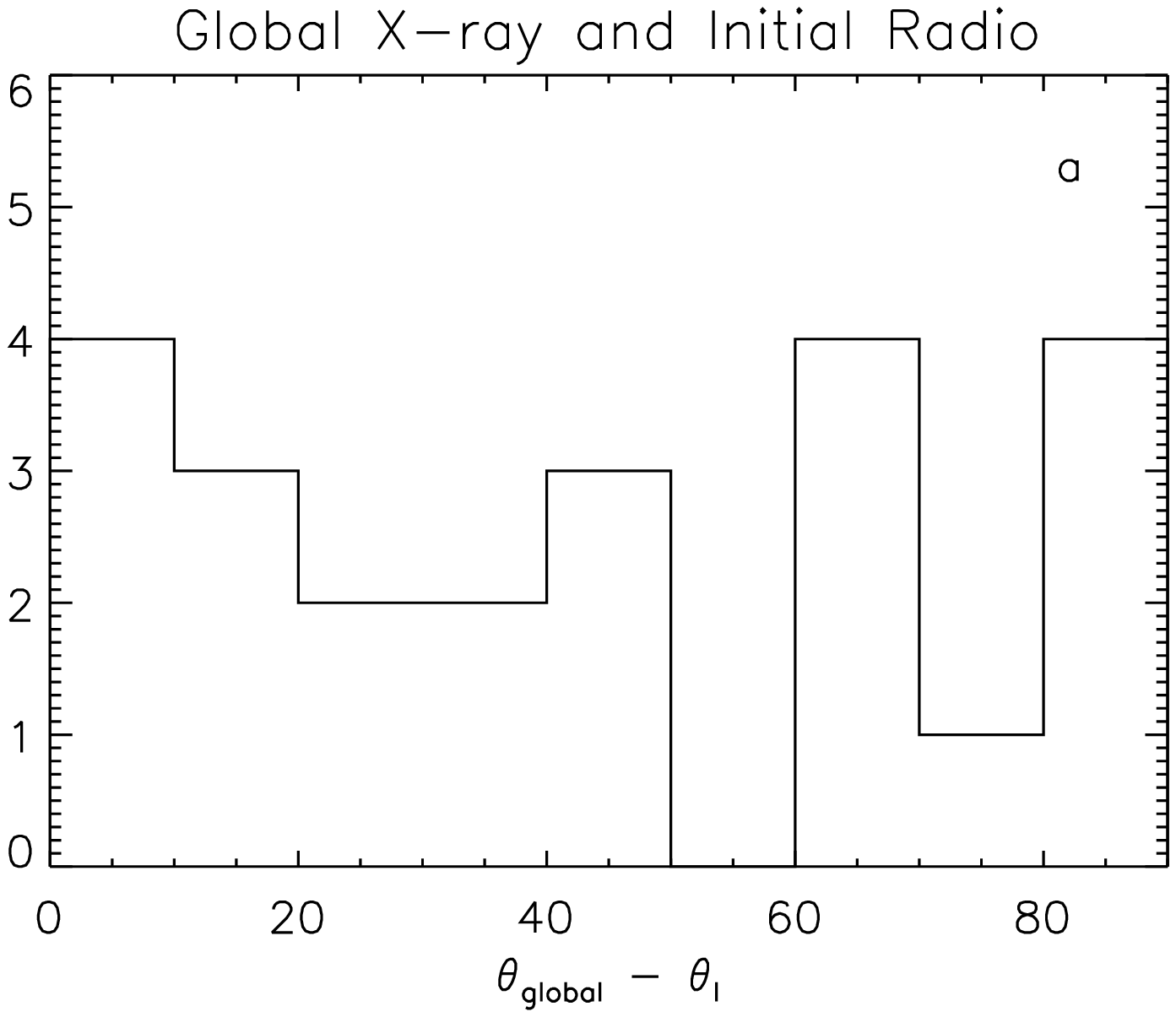,height=2.5in}
\psfig{figure=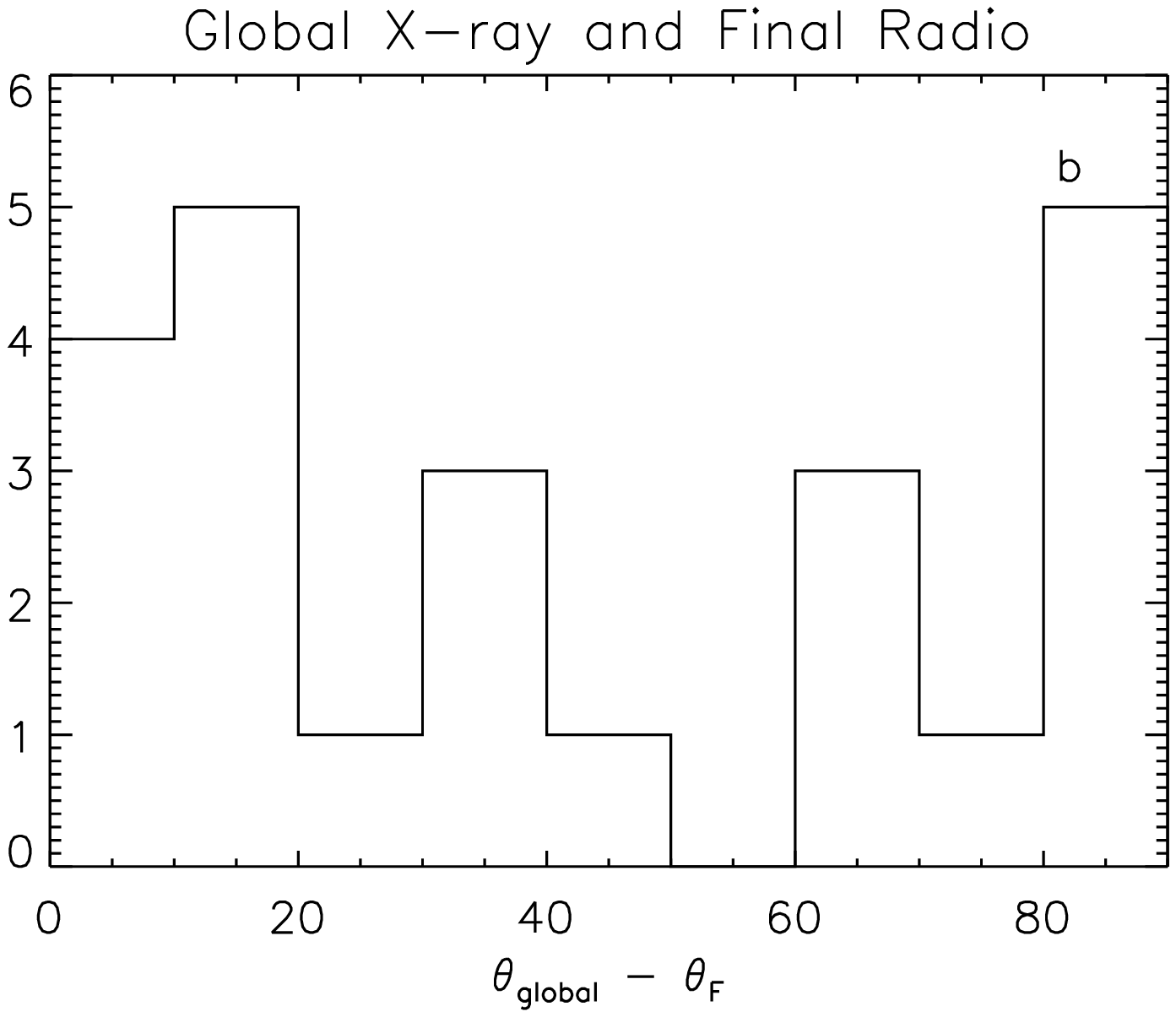,height=2.5in}}}
\centerline{\hbox{
\psfig{figure=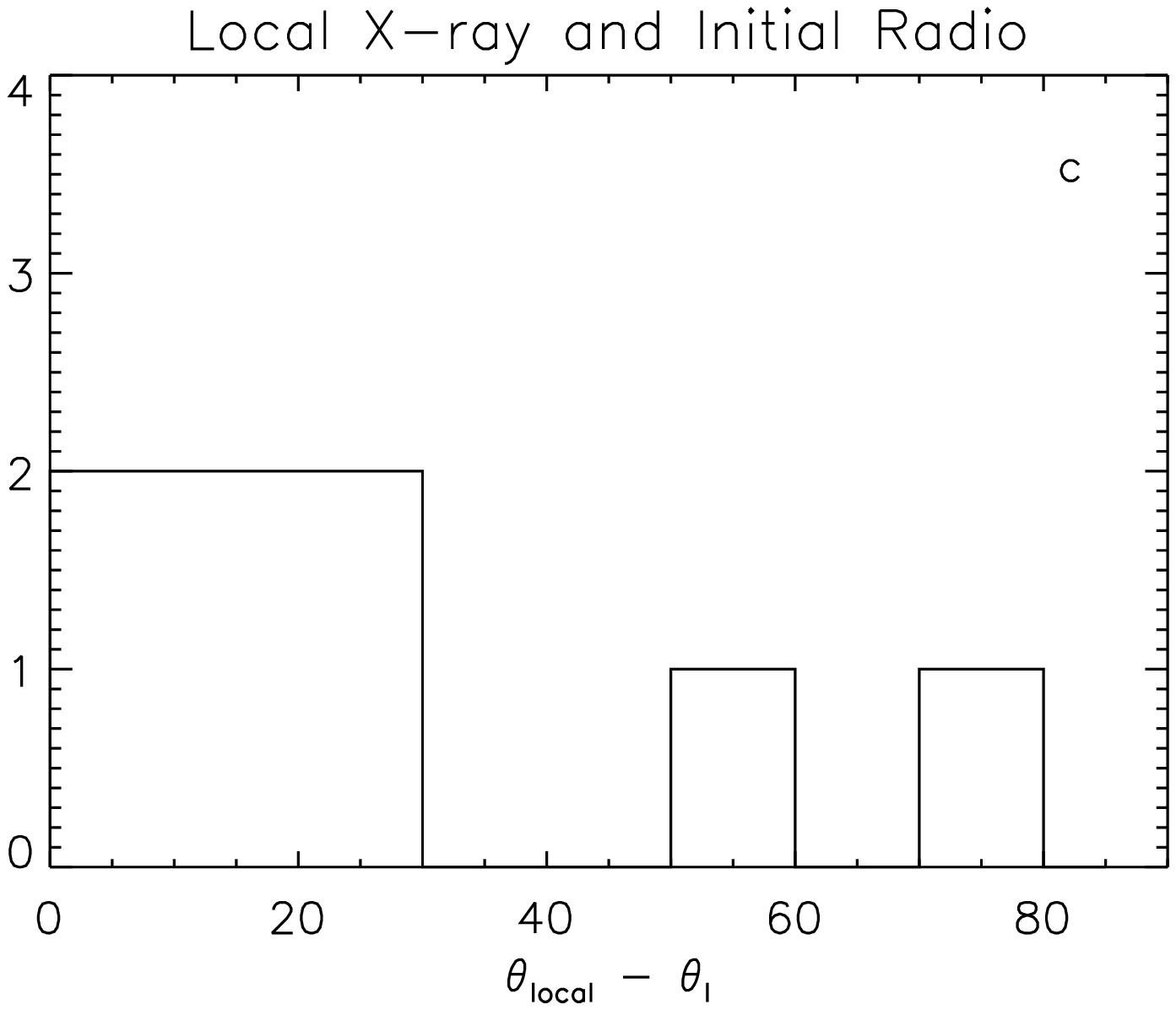,height=2.5in}
\psfig{figure=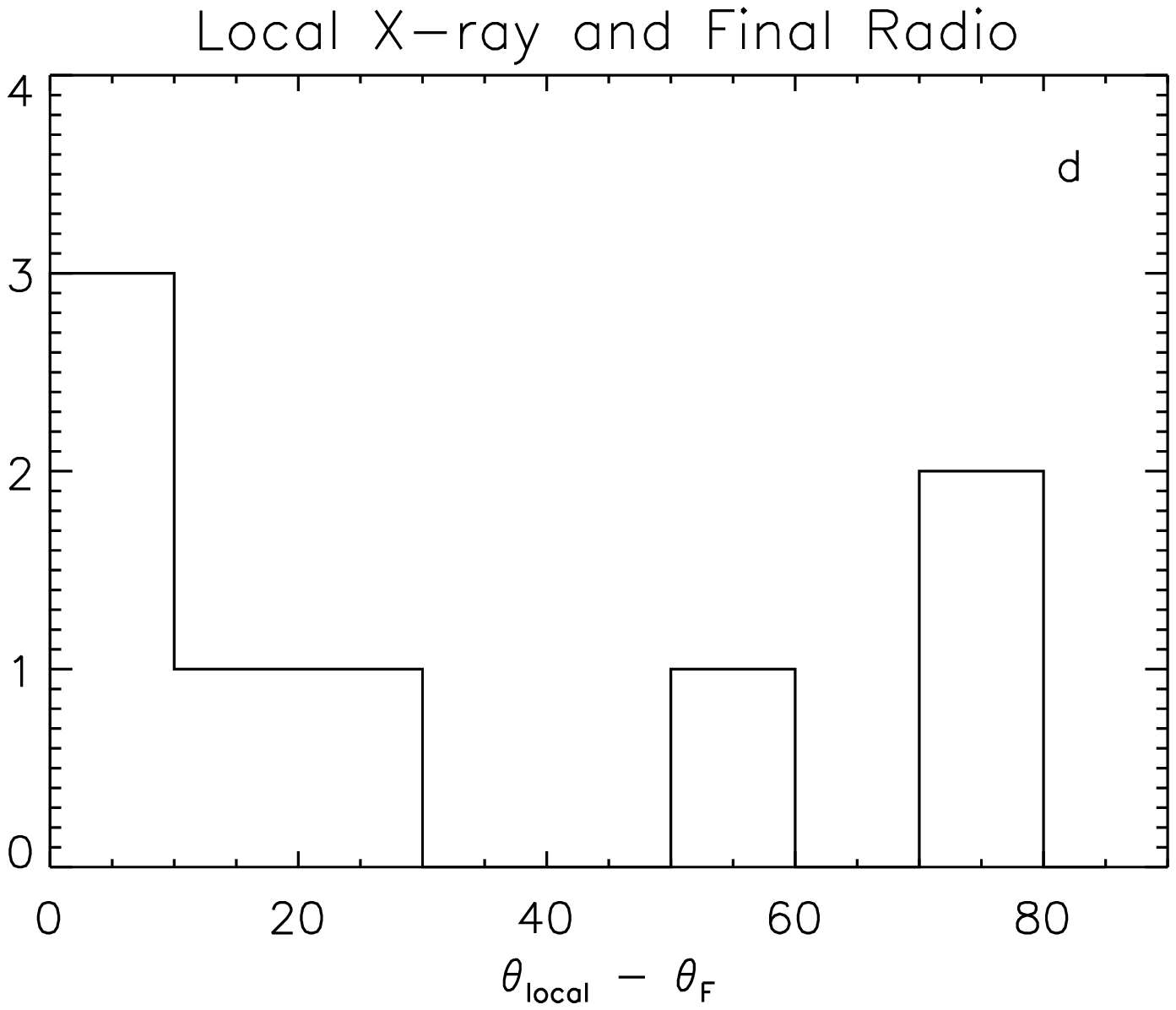,height=2.5in}}}
\label{fig9}
\caption{The difference between the various radio and X-ray
position angles, as described in the paper, folded into $\rm
90^{\circ}$.}
\end{figure*}

\subsection{Compact X-ray emission near NATs}
ER examined compact X-ray emission coincident with narrow-angle tailed
radio sources.  They found all such emission to be unresolved and
found a correlation between the X-ray flux and the core radio flux
density.  This led them to conclude that the X-ray emission associated
with NATs is most likely due to the AGN at the centres of the
galaxies.  We performed a similar analysis for our slightly different
sample of NAT galaxies. To best probe the X-ray emission associated
with the NATs, we subtracted an elliptical model of the overall
cluster emission from each image using the method described in \S 3.1.
The residual maps were examined for remaining compact X-ray emission
near the NAT radio sources.  X-ray detections with centroid positions
located more than 1.5 arcmin from the published centres of the radio
sources were discarded.  This left 40 per cent ($\frac{10}{25}$) of
the NATs with coincident, compact X-ray emission.  Radial profiles
were extracted from these regions in order to determine their extents.
Since the PSPC point spread function (PSF) is highly energy dependent,
the spectral information for these sources was needed. Thus we
extracted X-ray spectra for the regions using the {\sevensize PROS}
task QPSPEC and calculated the energy dependent PSF using the
{\sevensize FTOOLS} task PCRPSF.  The model of the PSF was compared to
the extracted radial profiles of the compact emission using the
$\chi^{\rm 2}$ statistic.  In agreement with the ER result, none of
the detected sources appear to be significantly resolved.

Only one NAT in our sample (3C264 in Abell 1367) had sufficient counts
to perform an adequate spectral fit to the X-ray emission.  The
spectral distribution was rebinned, so that all spectral channels had
at least 25 counts (S/N $\rm \sim 5$).  Using the {\sevensize XSPEC}
software package, we fit the spectrum separately with a power-law and
a Raymond \& Smith thermal model, to attempt to determine whether the
X-ray emission was due to an AGN or the galaxy ISM.  Similar to the ER
result, the spectral fits were inconclusive, giving equally good
statistics for both models.  Using {\it ASCA}, Forman (priv. comm.)
also found 3C264 to be well fit with either a 3 keV thermal model (the
same temperature as the surrounding emission) or a power-law.

The correlation between the core radio and the X-ray flux density
noted by ER, led them to suggest another constraint on unification
schemes between FR I radio sources and BL Lac objects.  We performed a
similar analysis for our sample, computing the X-ray and radio fluxes.
The X-ray emission was modeled as an absorbed power-law with energy
index 2.4 as done by ER.  The X-ray fluxes for all sources, including
upper limits on non-detections, are given in Table 5.  The
luminosities are also listed in Table 5 as well as limits on the
linear diameter of unresolved sources based on the approximate spatial
resolution of the PSPC.  Also included in the table are the 1.4 GHz
core radio power for the NATs. Whenever possible, the radio powers
were taken from O'Dea \& Owen (1985).  If the radio powers were not
available we measured them from the maps of the VLA Abell cluster
survey using {\sevensize AIPS}.  The X-ray luminosities and the core
radio powers for all 23 NATs are plotted in Fig. 10.

Since more than half the NATs (57 per cent) only have upper limits on
the X-ray flux, we used the survival analysis routines available in
the {\sevensize ASURV} software package Rev 1.1 (La Valley, Isobe, \&
Feigelson 1992), which implements the methods presented in Isobe,
Feigelson, \& Nelson (1986), to search for a correlation between core
radio power and X-ray luminosity.  Using the generalized Kendall's
$\tau$ correlation test, we find the probability for no correlation is
5.4 per cent, so although the data are suggestive of a correlation, we
cannot reject the null hypothesis.  With the current data, it is
impossible to determine the origin of the X-ray emission from NATs.

If the X-ray emission does not arise from an AGN, then what is its
origin?  One possibility is the galaxy ISM.  The JO model of NATs
proposes that there is an inner ISM with radius 4--50 kpc, not
stripped by ram pressure.  This gas is expected to have T $\sim$ 1 keV
(reflective of the stellar velocity dispersion) and X-ray emission
comparable to that observed in other ellipticals. Simulations of an
elliptical galaxy moving through the ICM (Balsara, Livio, \& O'Dea
1994) show that the outer galaxy ISM is strongly affected by ram
pressure stripping, producing shocks which propagate into the galaxy
core and which may also heat the ISM.  The current limit on the sizes
of the compact X-ray emission regions and the observed X-ray
luminosities (see Table 5) are consistent with this model and with
observations of other ellipticals (Fabbiano 1989).  Higher resolution
HRI X-ray observations of three NATs are currently underway which
should help us to discriminate between AGN and ISM models.

\section{NAT formation through cluster-subcluster mergers}
The current paradigm for NAT formation fails to account for the new
observed properties of NAT galaxies in rich clusters presented in this
paper.  The canonical view of NAT radio sources assumes the host
galaxies are moving at transonic velocities (i.e. 1--2 times the
velocity dispersion) through the dense ICM of spherical, relaxed
clusters.  In this picture the galaxy velocity and ICM density provide
the ram pressure necessary to bend the radio jets.  However, this
model does not account for the fact that NATs are preferentially found
in dynamically complex, possibly evolving clusters with significant
X-ray substructure.  Additionally, if the NAT tails are bent from ram
pressure by high velocity galaxies, then these radio sources should
show high peculiar motions.  However, NATs tend to have, on average,
velocities similar to a typical cluster galaxy.

We therefore propose a new model for NAT formation, in which NATs are
associated with dynamically complex clusters undergoing merger events.
Using an N-body + Hydro code, detailed simulations of merging clusters
have been performed (Roettiger, Burns \& Loken 1993, 1996; Schindler
\& M\"uller 1993).  One signature of merging is bulk flows in the
cluster gas with velocities $\rm > 1000 \ km \ s^{-1}$ aligned with
the X-ray elongation.  The resulting ram pressure of these ICM bulk
flows is sufficient to sweep back the radio jets (e.g. Loken et
al. 1995).  This ram pressure from the cluster gas would explain why
these radio sources, which tend to have typical cluster velocities,
exhibit the NAT radio morphology.  This is similar to our model for
the formation of wide-angle tailed radio sources (Gomez et al. 1997),
which also possess bent tails due to the motion of the ICM.  However,
since NAT radio sources tend to have lower radio powers (and possibly
lower jet momentum fluxes) than their WAT counterparts (O'Donoghue,
Eilek \& Owen 1993), this may explain why NAT jets are bent to larger
angles.
\begin{table*}
 \centering
 \begin{minipage}{140mm}
  \caption{NAT galaxy properties}
  \begin{tabular}{@{}cccccc@{}}
   Abell \# & IAU Name & X-ray Flux & X-ray Luminosity & Core Radio Power\footnote{O'Dea \& Owen (1985)} & X-ray Diameter \\
 & & $\rm 10^{-14} \ ergs \ cm^{-2} \ s^{-1}$ & $\rm 10^{41} \ ergs \
s^{-1}$ & $\rm 10^{23} \ W \
Hz^{-1}$ at 1.4 GHz & kpc \\[10pt] 
85 & 0039-095A & $\rm <6.8$ & $\rm <1.9$ & 0.20 & 29 \\
    & 0039-097 & $\rm 9.8\pm 1.5$ & $\rm 5.3\pm 0.8$ & 0.40 & \\
119 & 0053-015 & $\rm 3.0\pm 0.7$ & $\rm 1.0\pm 0.3$ & 1.58 & 24 \\
    & 0053-016 & $\rm <1.4$ & $\rm < 0.5$ & 0.33 & \\
194 & 0123-016A & $\rm 2.5\pm 0.4$ & $\rm 0.1\pm .02$ & 0.10 & 10 \\
496 & 0431-134 & $\rm 4.8\pm 1.1$ & $\rm 0.9\pm 0.2$ & 0.50 & 38 \\
514 & 0445-205 & $\rm <2.2$ & $\rm <2.0$ & 1.00 & 38 \\
    & 0446-205 & $\rm 16.6\pm 1.2$ & $\rm 15.4\pm 1.1$ & 3.16 & \\
754 & 0906-094 & $\rm 2.2\pm 0.9$ & $\rm 1.1\pm 0.4$ & 0.25 & 28 \\
1314 & 1131+493 & $\rm 14.4\pm 2.7$ & $\rm 3.0\pm 0.6$ & 5.01 & 18  \\
     & 1132+492 & $\rm 4.9\pm 1.7$ & $\rm 1.0\pm 0.3$ & 0.79 & \\
1367 & 1142+198 & $\rm 176.2\pm 3.6$ & $\rm 15.0\pm 0.3$ & 2.00 & 12 \\
     & 1141+202B & $\rm 26.6\pm 1.4$ & $\rm 2.3\pm 0.1$ & 0.03 & \\
1656 & 1256+281 & $\rm <4.6$ & $\rm <0.5$ & 0.02 & 13 \\
1775 & 1339+266B & $\rm <0.7$ & $\rm <0.6$ & 0.79 & 37 \\
1795 & 1346+268B & $\rm <1.5$ & $\rm <1.0$ & 1.00 & 32 \\
2142 & 1556+274 & $\rm <3.1$ & $\rm <4.2$ & 3.98 & 45 \\
2255 & 1712+640 & $\rm <2.2$ & $\rm <2.5$ & 0.79 & 41 \\      
     & 1712+641 & $\rm <2.2$ & $\rm <2.5$ & 1.58 & \\
2256 & 1705+786 & $\rm <1.6$ & $\rm <0.7$ & 0.16 & 31 \\
     & 1706+786 & $\rm <1.6$ & $\rm <0.7$ & 1.00 & \\
     & 1706+787 & $\rm <1.6$ & $\rm <0.7$ & 0.13 & \\
2589 & 2321+164 & $\rm <1.8$ & $\rm <0.6$ & 0.25 & 22 \\
\end{tabular}
\end{minipage}
\end{table*}

The ICM bulk motion may explain several other NAT observations as
well.  Numerical simulations of an elliptical galaxy moving
supersonically through the ICM (Balsara et al. 1994) show compression
of the ISM, with shock formation inside the galaxy, near the core.
Since free-free emission is proportional to the square of the gas
density, a small change in density produces a larger change in the
luminosity.  This may cause the gas in some NAT galaxies to become
detectable in the X-ray, and may explain the high fraction of NAT
galaxies (40 per cent) with unresolved X-ray emission.  Additionally,
the luminosities and sizes of the of the observed X-ray emission from
NATs (Table 5) is consistent with this model.

A second result of the Balsara, Livio, \& O'Dea simulation is the
periodic inflow of gas from the downstream side of the galaxy into the
galaxy core (inner $\sim$ 1 kpc), similar to the accretion of stellar
winds on to compact objects (Ishii et al. 1993; Ruffert \& Arnett
1994).  Although the resolution of the simulation is not high enough
to determine the fate of the gas at the core, some fraction of it
could fall on to the accretion disc.  This process could in turn feed
the central engine and increase the radio luminosity. These mechanisms
for enhanced radio and X-ray emission from NATs are an alternative
explanation for the correlation between core radio power and X-ray
luminosity discussed in \S 3.5.

Another common feature of NAT galaxies is the exceedingly long (up to
1 Mpc) radio tails (e.g. IC 711 in A1314, Vall\'ee \& Wilson 1976).
It is difficult to understand how the relativistic particles can
travel and then radiate at these distances from the cores (Ekers et
al. 1978; Wilson \& Vall\'ee 1977; Simon 1979; Vall\'ee \& Roger
1987).  Clearly, there must be a substantial particle reacceleration
mechanism.  But where does the energy come from to power the in-situ
particle acceleration?  The kinetic energy of the ICM from
cluster-subcluster mergers could provide the energy input necessary
for this reacceleration.
  
This new model for NAT formation is able to produce the morphology of
these unique radio sources.  In addition, it is able to account for
the X-ray substructure and velocity results presented in this paper,
and may explain the other NAT characteristics described above.
Therefore, we conclude that this new process for NAT formation
provides important insight into the cluster environment in which NATs
reside.  Since NATs are preferentially formed in merging clusters,
these radio sources can now be used as an additional diagnostic for
the dynamical state of clusters.  The presence of a NAT located in the
inner 0.3 $A_{\rm c}$ of a cluster may indicate a recent or ongoing
merger event.

\begin{figure}
\psfig{figure=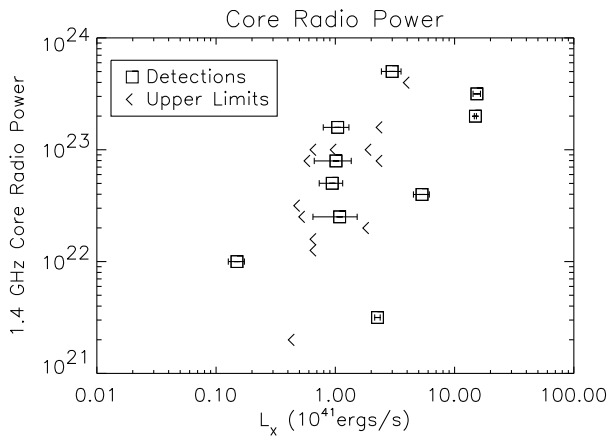,width=3.3in}
\label{fig10}
\caption{X-ray luminosity vs. 1.4GHz core radio power for
NATs.  The two parameters are not significantly correlated.}
\end{figure}

\section{Conclusions}
We have examined the X-ray and optical properties of nearby, rich
clusters of galaxies containing narrow-angle tailed radio sources.
This analysis has led us to some surprising conclusions.  First,
clusters containing NATs show a much higher degree of substructure in
their X-ray emission than their radio-quiet counterparts.  This
implies clusters containing NATs are mostly dynamically complex
systems, with possible recent or ongoing cluster-subcluster mergers.
Also, NAT galaxies within 0.3 $A_{\rm c}$ have a velocity distribution
very similar to that for other radio sources, and on average below the
threshold necessary to account for NAT jet bending due to ram
pressure.  These observations are inconsistent with the standard model
for the formation of NAT radio sources, which require a high velocity
galaxy traveling through the ICM to produce the ram pressure necessary
to bend the radio jets backward into a {\bf U}-shape.  Therefore, we
propose a new model for NAT formation in which NATs are, in general,
galaxies with small peculiar motions existing in dynamically complex
clusters which have undergone, or are undergoing, a merger with a
subcluster.  This merger scenario will create clusters with a high
degree of substructure in their X-ray surface brightness profiles.
Additionally, the ICM bulk motion caused by the merger will provide
the ram pressure necessary to bend the radio jets into the NAT
morphology.  The ICM ram pressure will also compress and shock the
ISM, possibly enhancing the compact X-ray emission detected in 40 per
cent of the NATs.  The luminosity and size of the compact emission
observed from NATs is consistent with this mechanism. The bulk motion
of the ICM may also provide the energy necessary for particle
reacceleration in NAT tails, as well as additional accretion material
for the central engine.  This new model of radio galaxies in a violent
ICM explains the observed morphological and kinematic properties of
NAT galaxies and their host clusters presented in this paper.

Some additional observations could help determine if the new NAT model
is correct.  First, our statistics for the correlations mentioned in
this paper are low because we must rely on archival PSPC data.  With
the launch of the {\it AXAF} X-ray satellite, the sample sizes can be
greatly extended.  Also, higher resolution X-ray images of the X-ray
bright NAT galaxies could allow us to determine the origin of the
X-ray emission.  At the redshifts of our sample, the {\it ROSAT} HRI
will probe the inner 3-5 kpc of the galaxy.  If the emission is still
point-like, the X-rays most likely originate from the AGN.  If a
component of the emission is resolved, however, it must originate from
some other mechanism. i.e. the galaxy ISM.  We have received time in
AO7 of the {\it ROSAT} mission to examine three NATs with the HRI.
This should help constrain the X-ray emission mechanism, and provide
additional insight into the formation of NAT radio sources.

\section*{Acknowledgments}

We thank Jane Turner at NASA GSFC for software help.  We also thank
Heinz Andernach, Alistair Edge, \& Chris O'Dea for useful discussions.
We acknowledge NRAO for its {\sevensize AIPS} software, NOAO for its
{\sevensize IRAF} software, and NASA GSFC for its {\sevensize FTOOLS}
and {\sevensize XSPEC} software.  This research was supported by NASA
grant NAGW-3152 to J.O.B. and F.N.O., and NSF grant AST-9896039 to J.O.B.

\section{Appendix:  individual sources}

\hskip 0.25 in {\bf Abell 85:} {\it Fig. 1a}.  This strong cooling
flow cluster has received much attention over the past few years in
both the optical and X-ray.  Analysis of the optical properties of
this cluster yield a virial mass of $\rm \approx 9\times 10^{14}$ $\rm
M_{\odot}$ and no obvious evidence of substructure (Girardi et
al. 1997; Malumuth et al. 1992).  However, the cD galaxy does have a
large peculiar velocity of 390 km $\rm s^{-1}$ with respect to the
cluster mean velocity.  The X-ray emission near the cD appears to be
inhomogeneous, with several clumps of X-ray enhancement which do not
seem to be associated with any faint galaxies in the optical
(Prestwich et al. 1995).  The southern-most NAT in the cluster is
associated with a clump of X-ray emission within $\rm \approx 100 \
kpc$ of the host radio galaxy.  This subclump may be the signature of
a possible accretion event.  Lima Neto et al. (1997) agree that the
southern X-ray extension contains a diffuse X-ray component in
addition to X-rays from individual galaxies.  High resolution radio
images (O'Dea \& Owen 1985) reveal that the radio source to the NE of
the cluster centre is also a NAT, although we do not see excess X-ray
emission associated with this source.

{\bf Abell 119:} {\it Fig. 1b}.  This cluster shows the best example
of NAT tails aligning with the asymmetric cluster X-ray emission.
Both NATs are pointing directly toward the NE elongation of the
cluster.  {\it HEAO 1} A2 spectral observations of A119 detected
possible nonisothermal emission from this cluster, which may be
evidence for a merger event (Henriksen 1993; Roettiger, Burns \&
Pinkney 1995).  {\it Einstein} IPC and HRI data also support the idea
of substructure in the X-ray morphology of the cluster, which closely
maps the asymmetric galaxy distribution (Fabricant et al. 1993).
However, examination of the optical position and velocity data for the
galaxies in A119 show no evidence of substructure (Girardi et
al. 1997; Kriessler \& Beers 1997; West \& Bothun 1990).  Polarization
studies of these two NATs show the magnetic fields to be oriented
along the jet, with high polarization in 0053-015, and much lower
polarization in 0053-016 (Mack, et al. 1993).

{\bf Abell 194:} {\it Fig. 1c}.  The striking feature in the X-ray
emission from this cluster is the strong elongation to the NW/SE,
where a similar feature is found in the galaxy distribution.  Some
galaxy position and velocity analyses do not show evidence of
substructure (Girardi et al. 1997; West \& Bothun 1990), while an
adaptive kernel analysis of the galaxy positions indicate the presence
of substructure (Kriessler \& Beers 1997).  Although the radio map of
this cluster is dominated by a twin jet to the NE, the NAT of interest
is the radio source near the centre of this map.  This cluster is also
the location of Minkowski's Object (Minkowski 1958), which is located
coincident with the NAT radio jet. It is suggested to be a cloud of
gas imbedded in the radio jet, which is now experiencing a burst of
star formation (Brodie, Bowyer \& McCarthy 1985; Van Breugel et al. 1985).

{\bf Abell 496:} {\it Fig. 1d}. This cooling flow cluster contains a
NAT in the SE coincident with a compact X-ray source.  The X-ray
emission near the cD galaxy in this cluster exhibits inhomogeneities
that may be associated with cooling flow filaments (Prestwich et
al. 1995).  As further evidence of a cooling flow, the emission line
nebula found at the centre of the cooling flow is attributed to thin
layers of self-absorbing gas around the surfaces of cold clouds
embedded in the hot, X-ray emitting gas (Crawford \& Fabian 1992).
Additionally, there is strong evidence for centrally enhanced metal
abundance from data gathered by the {\it Ginga} LAC and {\it Einstein}
SSS detectors (White et al. 1994).  A central cooling time of $T_{\rm
cool} \approx 2.1\pm0.3$ Gyr, and a mass inflow rate, {\it \.M} $\sim$
112 $\rm M_{\odot} \ yr^{-1}$ have been calculated (Edge, Stewart \&
Fabian 1992).  Malumuth et al. (1992) conclude that this cluster does
not show substructure in its galaxy velocity distribution, and
calculate a virial mass of $\rm \approx 7\times 10^{14}$ $\rm
M_{\odot}$.  However, other work (Kriessler \& Beers 1997; Mazure et
al. 1986) detects structure in the galaxy distribution, which is
mimicked in the slight elongation of the core X-ray emission to the
south, and the slight NW elongation of the outer X-ray emission.

{\bf Abell 514:} {\it Fig. 1e}.  This cluster possesses very clumpy,
asymmetric X-ray emission, with a large centroid shift in its
elliptical isophotes.  The two NATs are each associated with X-ray
subclumps in the cluster emission.

{\bf Abell 1775:} {\it Fig. 1f}.  The NAT in this cluster has its core
emission near the X-ray peak.  The tail extends in a direction
coincident with an X-ray enhancement to the NE of the X-ray centre.
This X-ray elongation is offset from the galaxy distribution which is
oriented mainly E-W (Trevese et al.  1992).  Galaxy velocity data show
this cluster to perhaps be two poor, interacting clusters, separated
by $\sim$3000 km $\rm s^{-1}$ (Oegerle, Hill \& Fitchett 1995).  The
NAT galaxy was once thought to be part of a bound binary galaxy pair
(Jenner 1974).  The measured velocity difference of these two galaxies
is $\rm \approx$ 1700 km $\rm s^{-1}$, and therefore, the galaxies
were assumed to be supermassive ($M > 2\times10^{13}$ $\rm M_{\odot}$)
in order for them to be bound (Chincarini et al. 1971).  However, the
measured dispersion of A1775 (1594 km $\rm s^{-1}$, Struble \& Rood
1991), allows for these galaxies to be unbound, and therefore of
average mass (Hintzen 1979).

{\bf Abell 754:} {\it Fig. 1g}.  A754 is an excellent example of a
cluster which has undergone a merger.  Henriksen \& Markevitch (1996)
found strong evidence for temperature structure in this cluster which
supports the earlier findings of Henry \& Briel (1996).  The area to
the NW of the cluster centre shows temperatures in excess of 12 keV,
while a cooler component below 5 keV is present to the SE.  The
elongation of the cluster X-ray centre may be due to an off-axis
merger of two subclusters (Roettiger, Stone \& Mushotzky 1998;
Henriksen \& Markevitch 1996).  The optical data are less conclusive,
with some studies showing evidence both for (Girardi et al. 1997;
Kriessler \& Beers 1997; West \& Bothun 1990; Dressler \& Schectman
1988) and against (Geller \& Beers 1982; Fabricant et al. 1986; Bird
1994) the presence of substructure.  This cluster exhibits an extreme
centroid shift and change in ellipticity in our analysis.  The NAT is
a very small radio source that was often mistaken for compact
emission, until viewed at high resolution.  Another NAT is coincident
with the knot of X-ray emission to the SW of the cluster, but its
location beyond 0.3 $A_{\rm c}$ placed it outside our sample selection
criteria.

{\bf Abell 1314:} {\it Fig. 1h}.  This cluster shows very clumpy,
elongated X-ray emission, and a strong X-ray centroid shift.  This
elongation is also present in the galaxy distribution, which shows a
definite ellipticity, and is oriented mainly E-W (Flin et al. 1995).
The cluster contains two narrow-angle tailed radio sources with the
western NAT's tail direction aligned with the local X-ray emission.
Although not detectable here, the NAT to the east (IC711) has one of
the longest tails ever detected, extending 17 arcmin (630 kpc) at a
wavelength of 74 cm (Vall\'ee \& Roger 1987).

{\bf Abell 1367:} {\it Fig. 1i}.  This cluster contains the strong
X-ray and radio source 3C264, located in the SW corner of the cluster.
A second NAT in the NW has often been misclassified as a double radio
source at low resolution.  Grebenev et al. (1995) performed a wavelet
analysis of the {\it ROSAT} PSPC and {\it Einstein} HRI images of this
cluster, and found sixteen extended X-ray features--nine of which had
galaxies coincident. This elongated NW region also contains several
spiral galaxies that appear to be infalling for the first time.  HI
maps of these spirals show the neutral hydrogen emission to be offset
from the optical centres and asymmetric in shape, implying the galaxy
discs are interacting with the ICM (Dickey \& Gavazzi 1991).  As
further evidence for ram pressure induced star formation, these spiral
galaxies all show giant HII regions aligned along a peripheral path,
suggestive of a bow shock formation mechanism (Gavazzi et al. 1995).
The luminosity function of spiral galaxies in this cluster deviates
from that of field spirals at the faint end, with bluer spirals
existing in the cluster.  This is another indication of galaxy-ICM
interactions causing enhanced star formation activity (Gavazzi,
Randone, \& Branchini 1995).

{\bf Abell 1656:} {\it Fig. 1j}.  The very rich Coma cluster has only
one NAT associated with it.  The tail of the NAT is oriented in a
direction similar to the core elongation.  It is believed that Coma
has recently undergone a cluster/subcluster merger (Burns et
al. 1994b).  This has been confirmed by several optical studies (Baier
1984; Fitchett \& Webster 1987; Mellier et al. 1988; Colless \& Dunn
1996; Zabludoff, Franx, \& Geller 1993; Caldwell et al. 1993);
however, the optical state of this cluster is still poorly understood
(e.g. Geller \& Beers 1982; Dressler \& Schectman 1988; West \& Bothun
1990).  The merger hypothesis is supported by filamentary X-ray
emission in the E and SE detected through a wavelet analysis
(Vikhlinin, Forman \& Jones 1997), and smaller X-ray subclumps
detected by the {\it Einstein} IPC (Davis \& Mushotzky 1993).  The
galaxy distribution in Coma is also peculiar, with the majority of
early-type galaxies lying along a filament to the NE/SW direction,
while the late-type galaxies are more symmetrically distributed (Doi
et al 1995).  The luminosity function in the core of Coma is well fit
by a power law with slope $\rm -1.42\pm.05$ over the range $
-19.4<M_{\rm R}<-11.4$ (Bernstein et al. 1995).  Caldwell et
al. (1993) detected a number of ``E+A'' (Dressler \& Gunn 1983)
galaxies, which are almost exclusively located to the SW of the
cluster centre.  This is a surprising result since most of these
post-starburst galaxies are detected in distant (z $\sim$ 0.2)
clusters (Butcher \& Oemler 1978, 1984; Dressler \& Gunn 1983; Couch
\& Sharples 1987).

{\bf Abell 1795:} {\it Fig. 1k}.  This cooling flow cluster shows
asymmetry in its galaxy distribution (Oegerle, Fitchett, \& Hoessel
1989).  However, combined with velocity information, no optical
substructure is present (Girardi et al. 1997).  The northern galaxy
elongation is mimicked by the X-ray emission which also shows an
excess to the north of the cluster core (Briel \& Henry 1996).  The
cooling flow is evidenced by a cooler temperature component to the
X-ray emission (Briel \& Henry 1996; White et al. 1994), with $T_{\rm
cool} \approx 2.1\pm0.3$ Gyr, and {\it \.M} $\rm \sim 478$ $\rm
M_{\odot} \ yr^{-1}$ (Edge et al. 1992).  Recently, attention has been
focused on {\it HST} examination of the cD galaxy.  This galaxy shows
excess $\rm H\alpha$ and UV filaments, as well as a dust lane
coincident with the edges of the radio lobes (Pinkney et al. 1996;
McNamara et al. 1996a, 1996b), which may imply massive star formation
in a cooling flow (Smith et al. 1997).

{\bf Abell 2255:} {\it Fig. 1l}. Multiwavelength studies indicate
A2255 to be a likely merger candidate (Burns et al. 1995).  The
velocity data for this cluster can either support (Tarenghi \& Scott
1976) or refute (Stauffer, Spinrad \& Sargent 1979) the merger
hypothesis.  The X-ray data show an elongation in the E-W direction,
and possible multiple temperature components (Feretti et al. 1997).
Additionally, Jones \& Forman (1984) measured a very large core radius
for this cluster ($\approx$ 0.4 Mpc).  A2255 contains a radio halo
source (Jaffe \& Rudnick 1979; Hanisch 1982), which may be produced as
a merger byproduct (Burns 1996) or by particle reacceleration by the
multiple NAT tails (Feretti et al. 1997).

{\bf Abell 2142:} {\it Fig. 1m}.  This cooling flow cluster has a
clear elongation in its X-ray emission to the NE/SW, with the NAT
oriented in roughly the same direction.  The cooling flow manifests
itself in a cool component fit to the X-ray spectrum, as well as
strong evidence for a centrally enhanced metal abundance (White et
al. 1994).  A2142 has a cooling time of $3.0\pm0.8$ Gyr, and a mass
inflow rate $\rm \sim 188$ $\rm M_{\odot} \ yr^{-1}$ Edge et
al. 1992).  X-ray surface brightness and temperature maps imply A2142
is in the late stages of a merger (Henry \& Briel 1996).  This cluster
also has a high velocity dispersion of 1280 km $\rm s^{-1}$ (Oegerle
et al. 1995).  The X-ray point source in the NE is most likely
associated with a Seyfert 1 cluster member galaxy.

{\bf Abell 2589:} {\it Fig. 1n}.  This cooling flow cluster shows an
obvious elongation to the N/S in the X-ray although the gas appears to
be isothermal (David, Jones \& Forman 1996).  There is also an
enhancement of emission to the south, near the NAT.  Optical data
analysed by Beers et al. (1991) found evidence for substructure.  They
also note that A2589 and the nearby cluster A2593 form a bound pair.
The envelope of the cD galaxy shows no evidence of a colour gradient,
which is consistent with the envelope being produced by mergers
(Mackie 1992).

{\bf Abell 2256:} {\it Fig. 1o}.  This cluster is one of the best
examples of clusters containing temperature substructure in the X-ray
emitting gas (Briel \& Henry 1994).  A wavelet analysis (Slezak,
Durrett \& Gerbal 1994), as well as isophotal analysis (Davis \&
Mushotzky 1993) of the X-ray emission shows extreme X-ray
substructure, in accord with the large centroid shift detected in this
analysis.  The X-ray emission from A2256 shows two distinct peaks,
which has been interpreted as the main cluster peak along with a
smaller subcluster in the process of merging (Briel et al. 1991;
Roettiger et al. 1995).  The main cluster has a temperature of 7.5 keV
(Hatsukade 1990), while the subcluster has a temperature ranging from
2 keV (Hatsukade 1990) to 4.6 keV (Miyaji et al. 1993) to 6.2 keV
(Markevitch 1996).  From {\it BBXRT} and {\it ROSAT} data, Miyaji et
al. (1993) calculate a gravitational mass of $\rm \sim
2.8-3.7\times10^{14}$ $\rm M_{\odot}$.  The galaxy density field
appears to be elliptical, and elongated in the same direction as the
cluster gas (Fabricant, Kent, \& Kurtz 1989). This cluster also
contains a radio halo source, several NAT sources, and a steep
spectrum radio source.  The halo source may actually be several NAT
galaxies which are distorted due to the subcluster merger
(R\"ottgering et al. 1994).

\label{lastpage}

\end{document}